\documentclass[twoside,11pt]{article}
\usepackage{jair, theapa, rawfonts}
\usepackage[small]{caption}
\usepackage[utf8]{inputenc} 
\usepackage[T1]{fontenc}    
\usepackage[hidelinks]{hyperref}       
\usepackage{url}            
\usepackage{booktabs}       
\usepackage{amsfonts}       
\usepackage{nicefrac}       
\usepackage{microtype}      
\usepackage{algorithm}
\usepackage{algpseudocode}
\usepackage{amsmath}
\usepackage{amssymb}
\usepackage{amsthm}
\usepackage{mathtools}
\usepackage{breakcites}
\usepackage{csquotes}
\usepackage{colortbl}
\usepackage{subcaption}
\usepackage[sort]{natbib}
\usepackage[shortlabels,inline]{enumitem}
\usepackage{arydshln}
\usepackage{mathabx}
\makeatletter
\newcommand{\inlineitem}[1][]{%
\ifnum\enit@type=\tw@
    {\descriptionlabel{#1}}
  \hspace{\labelsep}%
\else
  \ifnum\enit@type=\z@
       \refstepcounter{\@listctr}\fi
    \quad\@itemlabel\hspace{\labelsep}%
\fi}
\makeatother
\usepackage{thm-restate}
\usepackage{tabularx}
\usepackage{xcolor}
\usepackage{xspace}

\newcommand{\RR}{\mathbb{R}}

\newcommand{\GU}{GU\xspace}
\newcommand{\PU}{PU\xspace}
\newcommand{\GUs}{GU\xspace}
\newcommand{\PUs}{PU\xspace}
\newcommand{\GUl}{group unanimous\xspace}
\newcommand{\PUl}{pairwise unanimous\xspace}
\newcommand{\GUln}{group unanimity\xspace}
\newcommand{\PUln}{pairwise unanimity\xspace}

\newcommand{\his}{her\xspace}
\newcommand{\him}{her\xspace}
\newcommand{\he}{she\xspace}
\newcommand{\SP}{SP\xspace}
\newcommand{\SSP}{\SP}
\newcommand{\WSP}{WSP\xspace}

\newcommand{\WSPs}{WSP\xspace}
\newcommand{\SPlv}{strategyproof\xspace}
\newcommand{\SPl}{strategyproofness\xspace}
\newcommand{\SSPl}{\SPl}
\newcommand{\WSPl}{weak strategyproofness\xspace}
\newcommand{\WSPlv}{weakly strategyproof\xspace}
\newcommand{\minp}{\lambda}

\newcommand{\BiparCATAlgo}{\textsf{Divide-and-Rank}\xspace}

\newcommand{\defn}{:=}

\newcommand{\idf}{\mathrm{idf}}
\newcommand{\tf}{\mathrm{tf}}

\newcommand{\genreviewer}{r} 
\newcommand{\genpaper}{p} 
\newcommand{\cgenreviewer}{R}
\newcommand{\cgenpaper}{P}
\newcommand{\setreviewers}{{\cal \cgenreviewer}}
\newcommand{\setpapers}{{\cal \cgenpaper}}
\newcommand{\reviewer}[1]{\genreviewer_{#1}} 
\newcommand{\paper}[1]{\genpaper_{#1}} 
\newcommand{\indexreviewer}{i}
\newcommand{\indexpaper}{j}
\newcommand{\numreviewers}{m}
\newcommand{\numpapers}{n}
\newcommand{\setpaperselem}[1]{\setpapers_{#1}} 
\newcommand{\aggfunctionnotation}{f}
\newcommand{\aggfunction}[1]{\aggfunctionnotation(#1)} 
\newcommand{\revgraph}{\mathcal{G}} 
\newcommand{\conflictgraph}{\mathcal{C}}

\newcommand{\allpaperperms}[1]{\Pi(#1)} 
\newcommand{\paperrankingnotation}{\pi}
\newcommand{\paperranking}[1]{\paperrankingnotation(#1)} 
\newcommand{\indexedpaperranking}[2]{\paperrankingnotation^{(#2)}(#1)}
\newcommand{\indexedpaperrankshort}[1]{\paperrankingnotation^{(#1)}}
\newcommand{\paperrankingelem}[2]{\paperrankingnotation_{#2}(#1)} 
\newcommand{\higherrank}[2]{#1 \succ #2} 
\newcommand{\minreviewers}{\lambda}
\newcommand{\maxpapers}{\mu} 

\newcommand{\credible}{\textsf{Credible Subset}\xspace}

\newcommand{\dollarpartition}{\textsf{Dollar Partition}\xspace}
\newcommand{\exactdollarpartition}{\textsf{Exact Dollar Partition}\xspace}

\newcommand{\profile}{{\boldsymbol{\pi}}}

\newcommand{\profilerestrict}[1]{\profile_{#1}}

\newcommand{\pairopt}{\revgraph,\aggfunctionnotation}

\newcommand{\defeq}{\vcentcolon=}

\DeclarePairedDelimiterX{\inp}[2]{\langle}{\rangle}{#1, #2}

\newcommand{\gengraph}{\mathcal{H}} 

\usepackage{todonotes}

\theoremstyle{definition}
\newtheorem{theorem}{Theorem}[section]
\newtheorem{definition}[theorem]{Definition}

\newtheorem{lemma}[theorem]{Lemma}
\newtheorem{proposition}[theorem]{Proposition}
\newtheorem{corollary}[theorem]{Corollary}

\newcommand{\catname}{\textsf{Contract-and-Sort}\xspace}
\newcommand{\partition}{\textsf{Partition}\xspace}

\newcommand{\rcx}{\mathcal{R}_C}
\newcommand{\rcy}{\mathcal{R}_{\widebar{C}}}
\newcommand{\pcx}{\mathcal{P}_C}
\newcommand{\pcy}{\mathcal{P}_{\widebar{C}}}

\newcommand{\rankingx}{\pi_{C}}
\newcommand{\rankingy}{\pi_{\widebar{C}}}

\newcommand{\numberBeforePaperi}{k}
\newcommand{\indexrank}{p}
\newcommand{\indexrankk}{q}
\newcommand{\newfloorfunc}[1]{\left\lfloor #1 \right\rfloor}
\newcommand{\genassalgo}{\mathfrak{A}}
\newcommand{\genaggalgo}{\mathfrak{B}}
\newcommand{\truthfulranking}{\paperrankingnotation^*}
\newcommand{\estimatedranking}{\widehat{\paperrankingnotation}}
\newcommand{\constantnotation}{c}

\newcommand{\paperpos}{\ell}
\newcommand{\paperC}{\setpapers_C}

\newcommand{\randomvar}{X}

\jairheading{1}{1993}{1-15}{6/91}{9/91}
\ShortHeadings{On Strategyproof Conference Peer Review}
{Xu, Zhao, Shi, Zhang \& Shah}
\firstpageno{1}

\begin{document}

\title{On Strategyproof Conference Peer Review}

\author{\name Yichong Xu\thanks{Equal contribution.} \email yichongx@cs.cmu.edu \\
        \addr Machine Learning Department\\
        Carnegie Mellon University, Pittsburgh, PA, USA
        \AND
        \name Han Zhao\footnotemark[1] \email han.zhao@cs.cmu.edu \\
        \addr Machine Learning Department\\
        Carnegie Mellon University, Pittsburgh, PA, USA
        \AND
        \name Xiaofei Shi \email xiaofeis@andrew.cmu.edu \\
        \addr Department of Mathematical Sciences\\
        Carnegie Mellon University, Pittsburgh, PA, USA
        \AND
        \name Jeremy Zhang \email jbz@andrew.cmu.edu \\
        \addr Computer Science Department\\
        Carnegie Mellon University, Pittsburgh, PA, USA
        \AND
        \name Nihar B. Shah \email nihars@cs.cmu.edu \\
        \addr Machine Learning Department and Computer Science Department\\
        Carnegie Mellon University, Pittsburgh, PA, USA}


\maketitle

\begin{abstract}
We consider peer review in a conference setting where there is typically an overlap between the set of reviewers and the set of authors. This overlap can incentivize strategic reviews to influence the final ranking of one's own papers. In this work, we address this problem through the lens of social choice, and present a theoretical framework for strategyproof and efficient peer review. We first present and analyze an algorithm for reviewer-assignment and aggregation that guarantees strategyproofness and a natural efficiency property called unanimity, when the authorship graph satisfies a simple property. Our algorithm is based on the so-called partitioning method, and can be thought as a generalization of this method to conference peer review settings. We then empirically show that the requisite property on the authorship graph is indeed satisfied in the submission data from the ICLR conference, and further demonstrate a simple trick to make the partitioning method more practically appealing for conference peer review. Finally, we complement our positive results with negative theoretical results where we prove that under various ways of strengthening the requirements, it is impossible for any algorithm to be strategyproof and efficient.
\end{abstract}

\section{Introduction}
Peer review serves as an effective solution for quality evaluation in reviewing processes, especially in academic paper review~\citep{dorfler2017incentive,shah2017design} and massive open online courses (MOOCs)~\citep{diez2013peer,piech2013tuned,shah2013case}. However, despite its scalability, competitive peer review faces the serious challenge of being vulnerable to strategic manipulations~\citep{anderson2007perverse,thurner2011peer,alon2011sum,kurokawa2015impartial,kahng2017ranking}. By giving lower scores to competitive submissions, reviewers may be able to increase the chance that their own submissions get accepted. For instance, a recent experimental study~\citep{balietti2016peer} on peer review of art, published in the Proceedings of the National Academy of Sciences (USA), concludes
\begin{quote}\itshape
``...competition incentivizes reviewers to behave strategically,
which reduces the fairness of evaluations and the consensus
among referees.''
\end{quote}
As noted by~\citet{thurner2011peer}, even a small number of selfish, strategic reviewers can drastically reduce the quality of scientific standard. In the context of conference peer review,~\citet{langford201adversarialacad} calls academia inherently adversarial:
\begin{quote}\itshape
``It explains why your paper was rejected based on poor logic. The reviewer wasn’t concerned with research quality, but rather with rejecting a competitor.''
\end{quote}
Langford states that a number of people agree with this viewpoint. Thus the importance of peer review in academia and its considerable influence over the careers of researchers significantly underscores the need to design peer review systems that are insulated from strategic manipulations. 

In this work, we present a higher-level framework to address the problem of strategic behavior in conference peer review. We present an informal description of the framework here and formalize it later in the paper. The problem setting comprises a number of submitted papers and a number of reviewers. We are given a graph which we term as the ``conflict graph''. The conflict graph is a bipartite graph with the reviewers and papers as the two partitions of vertices, and an edge between any reviewer and paper if that reviewer has a conflict with that paper. Conflicts may arise due to authorship (the reviewer is an author of the paper) or other reasons such as being associated to the same institution. Given this conflict graph, there are two design steps in the peer review procedure: (i) assigning each paper to a subset of reviewers, and (ii) aggregating the reviews provided by the reviewers to give a final evaluation of each paper. Under our framework, the goal is to design these two steps of the peer-review procedure that satisfies two properties -- strategyproofness and efficiency.

Our goal is to design peer-review procedures that are strategyproof with respect to the given conflict graph. A peer-review procedure is said to be strategyproof if no reviewer can change the outcome for any paper(s) with which she/he has a conflict. This definition is formalized later in the paper. Strategyproofness not only reassures the authors that the review process is fair, but also ensures that the authors receive proper feedback for their work. We note that a strategyproof peer-review procedure alone is inadequate with respect to any practical requirements -- simply giving out a fixed, arbitrary evaluation makes the peer-review procedure strategyproof.  

Consequently, in addition to requiring strategyproofness, our framework measures the peer-review procedure with another yardstick -- that of efficiency. Informally, the efficiency of a peer-review procedure is a measurement of how well the final outcome reflects reviewers' assessments of the quality of the submissions, or a measurement of the accuracy in terms of the final acceptance decisions. There are several ways to define efficiency -- from a social choice perspective or a statistical perspective. In this paper, we consider efficiency in terms of the notion of unanimity in social choice theory: an agreement among all reviewers must be reflected in the final aggregation. 

In addition to the conceptual contribution based on this framework, we make several technical contributions towards this important problem. We first design a peer review algorithm which theoretically guarantees strategyproofness along with a notion of efficiency that we term ``group unanimity''. Our result requires only a mild assumption on the conflict graph of the peer-review design task. We show this assumption indeed holds true in practice via an empirical analysis of the submissions made to the  International Conference on Learning Representations (ICLR) conference\footnote{\url{https://openreview.net/group?id=ICLR.cc/2017/conference}}. Our algorithm is based on the popular partitioning method, and our positive results can be regarded as generalizing it to the setting of conference peer review. We further demonstrate a simple trick to make the partitioning method more practically appealing for conference peer review and validate it on the ICLR data.  

We then complement our positive results with negative results showing that one cannot expect to meet requirements that are much stronger than that provided by our algorithm. In particular, we show that under mild assumptions on the authorships, there is no algorithm that can be both strategyproof and ``pairwise unanimous''. Pairwise unanimity is a stronger notion of efficiency than group unanimity, and is also known as Pareto efficiency in the literature of social choice~\citep{brandt2016handbook}. We show that our negative result continues to hold even when the notion of strategyproofness is made extremely weak. We then provide a conjecture and insightful results on the impossibility when the assignment satisfies a simple ``connectivity'' condition. Finally, we connect back to the traditional settings in social choice theory, and show an impossibility when every reviewer reviews every paper. These negative results highlight the intrinsic hardness in designing strategyproof conference review systems.

\section{Related Work} 
As early as in the 1970s, Gibbard and Satterthwaite had already been aware of the importance of a healthy voting rule that is strategyproof in the setting of social choice~\citep{gibbard1973manipulation,satterthwaite1975strategy}. Nowadays, the fact that prominent peer review mechanisms such as the one used by the National Science Foundation~\citep{hazelrigg2013dear} and the one for time allocation on telescope~\citep{merrifield2009telescope} are manipulable has further called for strategyproof peer review mechanisms. 

Our work is most closely related to a series of works on strategyproof peer selection~\citep{de2008impartial,alon2011sum,holzman2013impartial,fischer2015optimal,kurokawa2015impartial,aziz2016strategyproof,kahng2017ranking,aziz2019strategyproof}, where agents cannot benefit from misreporting their preferences over other agents.\footnote{Some past literature refers to this requirement as ensuring that agents are ``impartial''. However, the term ``impartial'' also has connotations on (possibly implicit) biases due to extraneous factors such as some features about the agents~\citep{hojat2003impartial}. In this paper, we deliberately use the term ``strategyproof'' in order to make the scope of our contribution clear in that we do not address implicit biases.} \citet{de2008impartial} consider strategyproof decision making under the setting where a divisible resource is shared among a set of agents. Later, \citet{alon2011sum,holzman2013impartial} consider strategyproof peer approval voting where each agent nominates a subset of agents and the goal is to select one agent with large approvals. \citet{alon2011sum} propose a randomized strategyproof mechanism using partitioning that achieves provable approximate guarantee to the deterministic but non-strategyproof mechanism that simply selects the agent with maximum approvals. \citet{bousquet2014near} and \citet{fischer2015optimal} further extended and analyzed this mechanism to provide an optimal approximate ratio in expectation. Although the first partitioning-based mechanism partitions all the voters into two disjoint subsets, this has been recently extended to $k$-partition by \citet{kahng2017ranking}. In all these works, each agent is essentially required to evaluate \emph{all the other agents} except herself. This is impractical for conference peer review, where each reviewer only has limited time and energy to review a small subset of submissions. In light of such constraints, \citet{kurokawa2015impartial} propose an impartial mechanism (\credible) and provide associated approximation guarantees for a setting in which each agent is only required to review a few other agents. \credible~is a randomized mechanism that outputs a subset of $k$ agents, but it has non-zero probability returns an empty set. Based on the work of~\citet{de2008impartial}, \cite{aziz2016strategyproof} propose a mechanism for peer selection, termed as \dollarpartition, which is strategyproof and satisfies a natural monotonicity property. Empirically the authors showed that \dollarpartition outperforms \credible consistently and in the worst case is better than partition-based approach. However, even if the target output size is $k$, \dollarpartition may return a subset of size strictly larger than $k$. This problem has recently been fixed by the \exactdollarpartition mechanism~\citep{aziz2019strategyproof}, which empirically selects more high-quality agents more often and consistently than \credible. Our positive results, specifically our \BiparCATAlgo algorithm presented subsequently, borrows heavily from this line of literature. That said, our work addresses the application of conference peer review which is more general and challenging as compared to the settings considered in past works.

Our setting of conference peer review is more challenging as compared to these past works as each reviewer may author multiple papers and moreover each paper may have multiple authors as reviewers. Specifically, the conflict graph under conference peer review is a general bipartite graph, where conflicts between reviewers and papers can arise not only because of authorships, but also advisor-advisee relationships, institutional conflicts, etc. In contrast, past works focus on applications of peer-grading and grant proposal review, and hence consider only one-to-one conflict graphs (that is, where every reviewer is conflicted with exactly one paper).  

Apart from the most important difference mentioned above, there are a couple of other differences of this work as compared to some past works. In this paper we focus on ordinal preferences where each reviewer is asked to give a total ranking of the assigned papers, as opposed to providing numeric ratings. We do so inspired by past literature~\citep{barnett2003modern,stewart2005absolute,douceur2009paper,tsukida2011analyze,shah2013case,shah2016estimation} which highlights the benefits of ordinal data in terms of avoiding biases as well as allowing for a more direct comparison between papers. Secondly,  while most previous mechanisms either output a single paper or a subset of papers, we require our mechanism to output a total ranking over all papers. We consider this requirement since this automated output in practice will be used by the program chairs as a guideline to make their decisions, and this more nuanced data comprising the ranking of the papers can be more useful towards this goal.

A number of papers study various other aspects of conference peer review, and we mention the most relevant ones here. Several works~\citep{Hartvigsen99assignment,charlin13tpms,Garg2010papers,stelmakh2018assignment} design algorithms for assigning reviewers to papers under various objectives, and these objectives and algorithms may in fact be used as alternative definitions of the objective of ``efficiency'' studied in the present paper. The papers~\citet{roos2011calibrate,ge13bias,wang2018your} consider review settings where reviewers provide scores to each paper, with the aim of addressing the problems of miscalibration in these scores. \cite{tomkins2017reviewer,stelmakh2019testing} study biases in peer review, \cite{noothigattu2018choosing} address issues of subjectivity,~\cite{gao2019does} investigate rebuttals, and~\cite{fiez2019super} improve the efficiency of the bidding process. Experiments and empirical evaluations of conference peer reviews can be found in~\citet{nips14experiment,claire2008soda, connolly2014longitudinal,shah2017design,tomkins2017reviewer,noothigattu2018choosing,gao2019does}.


\section{Problem setting}
\label{sec:preliminary}

In this section, we first give a brief introduction to the setting of our problem, and then introduce the notation used in the paper. We then formally define various concepts and properties to be discussed in the subsequent sections. 

Modern review process is governed by four key steps: (i) a number of papers are submitted for review; (ii) each paper is assigned to a set of reviewers; (iii) reviewers provide their feedback on the papers they are reviewing; and (iv) the feedback from all reviewers is aggregated to make final decisions on the papers. Let $\numreviewers$ be the number of reviewers and $\numpapers$ be the number of submitted papers. Define $\setreviewers \defn \{\reviewer{1}, \ldots, \reviewer{\numreviewers}\}$ to be the set of $\numreviewers$ reviewers and $\setpapers \defn \{\paper{1}, \ldots, \paper{\numpapers}\}$ to be the set of $\numpapers$ submitted papers.

The review process must deal with conflicts of interest. To characterize conflicts of interest, we use a bipartite graph $\conflictgraph$ with vertices $(\setreviewers, \setpapers)$, where an edge is connected between a reviewer $\genreviewer$ and a paper $\genpaper$ if there exists some conflict of interests between reviewer $\genreviewer$ and paper $\genpaper$.  Reviewers who do not have conflicts of interest with any paper are nodes with no edges. Given the set of submitted papers and reviewers, this graph is fixed and cannot be controlled. Note that the conflict graph $\conflictgraph$ defined above can be viewed as a generalization of the authorship graph in the previously-studied settings~\citep{merrifield2009telescope,alon2011sum,holzman2013impartial,fischer2015optimal,kurokawa2015impartial,aziz2016strategyproof,kahng2017ranking} of peer grading and grant proposal review, where each reviewer (paper) is connected to at most one paper (reviewer).

The review process is modeled by a second bipartite graph $\revgraph$, termed as \emph{review graph}, that also has the reviewers and papers $(\setreviewers, \setpapers)$ as its vertices. This review graph has an edge between a reviewer and a paper if that reviewer reviews that paper. For every reviewer $\reviewer{\indexreviewer}~(\indexreviewer \in [\numreviewers])$,\footnote{We use the standard notation $[\kappa]$ to represent the set $\{1,\ldots,\kappa\}$ for any positive integer $\kappa$.} we let $\setpaperselem{\indexreviewer} \subseteq \setpapers$ denote the set of papers assigned to this reviewer for review, or in other words, the neighborhood of node $\reviewer{\indexreviewer}$ in the bipartite graph $\revgraph$. The program chairs of the conference are free to choose this graph, but subject to certain constraints and preferences. To ensure balanced workloads across reviewers, we require that every reviewer is assigned at most $\maxpapers$ papers for some integers $1 \leq \maxpapers \leq \numpapers$. In other words, every node in $\setreviewers$ has at most $\maxpapers$ neighbors (in $\setpapers$) in graph  $\revgraph$. Additionally, each paper must be reviewed by a certain minimum number of reviewers, and we denote this minimum number as $\minreviewers$. Thus every node in the set $\setpapers$ must have at least $\minreviewers$ neighbors (in $\setreviewers$) in the graph $\revgraph$. For any (directed or undirected) graph $\gengraph$, we let the notation $E_\gengraph$ denote the set of (directed or undirected, respectively) edges in graph $\gengraph$. 

At the end of the reviewing period, each reviewer provides a total ranking of the papers that she/he reviewed. For any set of papers $\setpapers' \subseteq \setpapers$, we let $\allpaperperms{\setpapers'}$ denote the set of all permutations of papers in $\setpapers'$. Furthermore, for any paper $\paper{\indexpaper} \in \setpapers'$ and any permutation $\paperranking{\setpapers'} \in \allpaperperms{\setpapers'}$, we let $\paperrankingelem{\setpapers'}{\indexpaper}$ denote the position of paper $\paper{\indexpaper}$ in the permutation $\paperranking{\setpapers'}$. At the end of the reviewing period, each reviewer $\reviewer{\indexreviewer}~(\indexreviewer \in [\numreviewers])$ submits a total ranking $\indexedpaperranking{\setpaperselem{\indexreviewer}}{\indexreviewer} \in \allpaperperms{\setpaperselem{\indexreviewer}}$ of the papers in $\setpaperselem{\indexreviewer}$. We define a (partial) ranking profile $\profile \defn (\indexedpaperranking{\setpaperselem{1}}{1}, \ldots, \indexedpaperranking{\setpaperselem{\numreviewers}}{\numreviewers} )$ as the collection of rankings from all the reviewers. When the assignment $\setpaperselem{1},\ldots,\setpaperselem{\numreviewers}$ of papers to reviewers is fixed, we use the shorthand $(\indexedpaperrankshort{1},\ldots,\indexedpaperrankshort{\numreviewers})$ for profile $\profile$. For any subset of papers $\setpapers' \subseteq \setpapers$, we let $\profilerestrict{\setpapers'}$ denote the restriction of $\profile$ to only the induced rankings on $\setpapers'$. Finally, when the ranking under consideration is clear from context, we use the notation $\higherrank{\genpaper}{\genpaper'}$ to say that paper $\genpaper$ is ranked higher than paper $\genpaper'$ in the ranking. 

Under this framework, the goal is to jointly design: (a) a paper-reviewer assignment scheme, that is, edges of the graph $\revgraph$, and (b) an associated review aggregation rule $\aggfunctionnotation:\prod_{\indexreviewer=1}^{\numreviewers} \allpaperperms{\setpaperselem{\indexreviewer}} \to \allpaperperms{\setpapers}$ which maps from the ranking profile to an aggregate total ranking of all papers.\footnote{To be clear, the function $\aggfunctionnotation$ is tied to the assignment graph $\revgraph$. The graph $\revgraph$ specifies the sets $(\setpaperselem{1},\ldots,\setpaperselem{\numreviewers})$, and then the function $\aggfunctionnotation$ takes permutations of these sets of papers as its inputs. We omit this from the notation for brevity.} For any aggregation function $\aggfunctionnotation$, we let $\aggfunctionnotation_\indexpaper(\profile)$ be the position of paper $\paper{j}$ when the input to $\aggfunctionnotation$ is the profile $\profile$.

We note that although we assume ordinal feedback from the reviewers, our results continue to hold if we have review scores as our input instead of rankings; our framework is flexible enough to take the scores into account (cf. Section \ref{sec:bicat_algo}).

In what follows we define strategyproofness and efficiency that any conference review mechanism $\aggfunctionnotation$ should satisfy under our paper-review setting. Inspired by the theory of social choice, in this paper we define the notion of efficiency via two variants of  {``unanimity''}, and we also discuss two natural notions of strategyproofness. 

    \subsection{Strategyproofness}
Intuitively, strategyproofness means that a reviewer cannot benefit from being dishonest. In the context of conference review, strategyproofness is defined with respect to a given conflict graph $\conflictgraph$; we recall the notation $E_{\conflictgraph}$ as the set of edges of graph $\conflictgraph$. It means that a reviewer cannot change the position of \his conflicting papers, by manipulating the ranking she provides.
\begin{definition}[Strategyproofness, \SSP]\label{def:s-strategyproof}
A review process $(\pairopt)$ is called {strategyproof} with respect to a conflict graph $\conflictgraph$ if for every reviewer $\reviewer{\indexreviewer}\in \setreviewers$ and paper $\paper{\indexpaper} \in \setpapers$ such that $(\reviewer{\indexreviewer}, \paper{\indexpaper})\in E_{\conflictgraph}$ the following condition holds: for every pair of  profiles (under assignment $\revgraph$) that differ only in the ranking given by reviewer $\reviewer{\indexreviewer}$, the position of $\paper{\indexpaper}$ is unchanged.\footnote{A related (and weaker) definition of strategyproof is that the position of any $\paper{\indexpaper}$ cannot be \emph{improved}. It is easy to show that any mechanism that satisfies the weaker notion can also satisfy our notion of strategyproofness.} Formally, $\forall \profile = (\indexedpaperrankshort{1},\ldots,\indexedpaperrankshort{i-1},\indexedpaperrankshort{i},\indexedpaperrankshort{i+1},\ldots,\indexedpaperrankshort{\numreviewers})$ and $\profile' = (\indexedpaperrankshort{1},\ldots,\indexedpaperrankshort{i-1},{\indexedpaperrankshort{i}}',\indexedpaperrankshort{i+1},\ldots,\indexedpaperrankshort{\numreviewers})$, it must be that $\aggfunctionnotation_{\indexpaper}(\profile) = \aggfunctionnotation_{\indexpaper}(\profile')$.
\end{definition}

A strategyproof peer review procedure alone is inadequate with respect to any practical requirements -- simply giving out a fixed, arbitrary evaluation makes the peer review procedure strategyproof. We therefore consider efficiency of the procedure in the next section, to ensure that the authors receive meaningful and helpful feedback for their work.

\subsection{Efficiency (unanimity)}

Consequently, in addition to requiring strategyproofness, we measure the peer review procedure with another yardstick -- efficiency. The peer review procedure needs to not only reassure the authors that the review process is fair, but also ensure that the authors receive proper feedback for their work in an efficient way.

In this work, we consider efficiency of a peer-review process in terms of the notion of unanimity. Unanimity is one of the most prevalent and classic properties to measure the efficiency of a voting system in the theory of social choice~\citep{fishburn2015theory}. At a colloquial level, unanimity states that when there is a common agreement among all reviewers, then the aggregation of their opinions must also respect this agreement. In this paper we discuss two kinds of unanimity, termed group unanimity (GU) and pairwise unanimity (PU). Both kinds of unanimity impose requirements on the aggregation function for any given reviewer assignment. 

We first define group unanimity: 
\begin{definition}[Group Unanimity, \GU]
\label{def:group}
We define $(\pairopt)$ to be group unanimous (\GU) if the following condition holds for every possible profile $\profile$. If there is a non-empty set of papers $\setpapers' \subset \setpapers$ such that every reviewer ranks the papers \he reviewed from $\setpapers'$ higher than those \he reviewed from $\setpapers\setminus \setpapers'$, then $\aggfunction{\profile}$ must have $\paper{x} \succ \paper{y}$ for every pair of papers $\paper{x} \in \setpapers'$ and $\paper{y} \in \setpapers \setminus \setpapers'$ such that at least one reviewer has reviewed both $\paper{x}$ and $\paper{y}$. 
\end{definition}
Intuitively, group unanimity says that if papers can be partitioned into two sets such that every reviewer who has reviewed papers from both sets agrees that the papers \he has reviewed from the first set are better than what \he reviewed from the second set, then the final output ranking should respect this agreement. 

Our second notion of unanimity, termed pairwise unanimity, is a local refinement of group unanimity. This notion is identical to the classical notion of unanimity stated in Arrow's impossibility theorem~\citep{arrow1950difficulty} -- the classical unanimity considers every reviewer to review all papers (that is, $\setpapers_i = \setpapers, \forall i\in[m]$), whereas our notion is also defined for settings where reviewers may review only subsets of papers. 
\begin{definition}[Pairwise Unanimity, \PU]
\label{def:p-unanimity}
We define $(\pairopt)$ to be pairwise unanimous (\PU) if the following condition holds for every possible profile $\profile$ and every pair of papers $\paper{\indexpaper_1},\paper{\indexpaper_2}\in \setpapers$: 
If at least one reviewer has reviewed both $\paper{\indexpaper_1}$ and $\paper{\indexpaper_2}$ and all the reviewers that have reviewed $\paper{\indexpaper_1}$ and $\paper{\indexpaper_2}$ agree on $\paper{\indexpaper_1} \succ \paper{\indexpaper_2}$, then $\aggfunctionnotation_{\indexpaper_1}(\profile) \succ  \aggfunctionnotation_{\indexpaper_2}(\profile)$. 
\end{definition}

An important property is that \PUln is stronger than \GUln.
\begin{proposition}
\label{lemma:unanimity}
    If $(\revgraph,\aggfunctionnotation)$ is \PUl, then $(\pairopt)$ is also \GUl.
\end{proposition}
The proof of this proposition is provided in Section \ref{sec:proof:lemma:unanimity}. 

\section{Positive Theoretical Results and Algorithm}
\label{sec:pos}

In this section we consider the design of reviewer assignments and aggregation rules for strategyproofness and group unanimity (efficiency). It is not hard to see that strategyproofness and group unanimity cannot be simultaneously guaranteed for arbitrary conflict graphs $\conflictgraph$, for instance, when $\conflictgraph$ is a fully-connected bipartite graph. Prior works on this topic consider a specific class of conflict graphs --- those with one-to-one relations between papers and reviewers --- which do not capture conference peer review settings. We consider a more general class of conflict graphs and present an algorithm based on the partitioning-based method~\citep{alon2011sum},  which we show can achieve \GUl and \SPl.

We then empirically demonstrate, using submission data from the ICLR  conference, that this class of conflict graphs is indeed representative of peer review settings. We observe that the quality of the reviewer assignment under our method (that guarantees strategyproofness) is only slightly lower as compared to the optimal quality in the absence of strategyproofing requirements. Finally, we present a simple trick to significantly improve the practical appeal of our algorithm (and more generally the partitioning method) to conference peer review.

\subsection{\label{sec:bicat_algo} The \BiparCATAlgo~Algorithm}

We now present our ``\BiparCATAlgo'' framework consisting of the reviewer assignment algorithm (Algorithm \ref{alg:assign}) and the rank aggregation algorithm (Algorithm \ref{alg:aggregate}). 
At a high level, our algorithm performs a partition of the reviewers and papers for assignment, and aggregates the reviews by computing a ranking which is consistent with any group agreements. The \BiparCATAlgo algorithm works for a general conflict graph $\conflictgraph$ as long as the conflict graph can be divided into two reasonably-sized disconnected components. 

Importantly, the framework is simple yet flexible in that the assignment within each partition and the aggregation among certain groups of papers can leverage any existing algorithm for assignment and aggregation respectively, which is useful as it allows to further optimize various other metrics in addition to strategyproofness and unanimity. 

Below we describe our framework in more detail. We first introduce the assignment procedure in Algorithm~\ref{alg:assign}.

\begin{algorithm}[htb]
\centering
\caption{\BiparCATAlgo assignment}
\label{alg:assign}
\begin{algorithmic}[1]
\Require conflict graph $\conflictgraph$, parameters $\minreviewers, \maxpapers$, assignment algorithm $\genassalgo$
\Ensure an assignment of reviewers to papers
\State $(\rcx, \pcx)$, $(\rcy, \pcy)\leftarrow\partition(\conflictgraph, \minreviewers, \maxpapers)$
\State use algorithm $\genassalgo$ to assign papers $\pcy$ to reviewers $\rcx$ 
\State use algorithm $\genassalgo$ to assign papers $\pcx$ to reviewers $\rcy$
\State \textbf{return} the union of assignments from step 2 and 3
\\\hrulefill
\Procedure{\partition}{conflict graph $\conflictgraph$, parameters $\minreviewers, \maxpapers$}
\State  run a BFS on $\conflictgraph$ to get connected $K$ components $\{(\setreviewers_k, \setpapers_k)\}_{k=1}^K$
\State  let $r_k = |\setreviewers_k|, p_k = |\setpapers_k|$, $\forall k\in[K]$
\State  initialize a table $T[\cdot, \cdot, \cdot]\in\{0, 1\}^{K\times (\numreviewers + 1)\times (\numpapers + 1)}$ so that $T[1, r_1, p_1] = T[1, 0, 0] = 1$, otherwise 0
\For {$k = 2$ to $K$}
    \State  $T[k, r, p] = T[k-1, r, p] \vee T[k-1, r-r_k, p-p_k],~\forall 0\leq r\leq \numreviewers, 0\leq p\leq \numpapers$ 
\EndFor
\State  for $0\leq r\leq \numreviewers, 0\leq p\leq \numpapers$, if there is no $T[K, r, p] = 1$ such that $\max\{\frac{p}{\numreviewers - r}, \frac{\numpapers - p}{r}\}\leq \frac{\maxpapers}{\minp}$, return \textsc{error}
\State  use the standard backtracking in the table $T[\cdot, \cdot, \cdot]$ to return $(\rcx, \pcx)$ and $(\rcy, \pcy)$
\EndProcedure
\end{algorithmic}
\end{algorithm}

The \BiparCATAlgo assignment algorithm begins by partitioning the conflict graph into two disconnected components such that (1) they meet the requirements specified by $\maxpapers$ and $\minreviewers$; and (2) the two disconnected components have roughly equal size in terms of number of nodes. This is achieved using the subroutine \partition. In more detail, \partition first runs a breadth-first-search (BFS) algorithm to partition the original conflict graph into $K$ connected components, where the $k$th connected component contains $r_k\geq 0$ reviewers and $p_k \geq 0$ papers. Next, the algorithm performs a dynamic programming to compute all the possible subset sums, i.e., sum of the number of reviewers and the number of papers in a given subset, achievable by the $K$ connected components. Here $T[k,r,p]=1$ means that there exists a partition of the first $k$ components such that one side of the partition has $r$ reviewers and $p$ papers, and 0 otherwise. The last step is to check whether there exists a subset $C$ satisfying the constraint given by $\minreviewers$ and $\maxpapers$, and if so, runs a standard backtracking algorithm along the table to find the actual subset $C$. Clearly the \partition runs in $O(K\numpapers\numreviewers)$, and since $K\leq \numpapers\numreviewers$, it runs in polynomial time in the size of the input conflict graph $\conflictgraph$.

In the next step, the algorithm assigns papers to reviewers in a fashion that guarantees each paper is going to be reviewed by at least $\minreviewers$ reviewers and each reviewer reviews at most $\maxpapers$ papers. The assignment of papers in any individual component (to reviewers in the other component) can be done using any assignment algorithm (taken as an input $\genassalgo$) as long as the algorithm can satisfy the $(\maxpapers, \minreviewers)$-requirements. Possible choices for the algorithm $\genassalgo$ include the popular Toronto paper matching system~\citep{charlin13tpms} and  others~\citep{Hartvigsen99assignment,Garg2010papers,stelmakh2018assignment}. We can also use the typical reviewer bidding system, while constraining the reviewers in $\rcx$ to review $\pcy$ and $\rcy$ to review $\pcx$.

We then introduce to the aggregation procedure in Algorithm~\ref{alg:aggregate}.
\begin{algorithm}[htb]
\centering
\caption{\BiparCATAlgo aggregation}
\label{alg:aggregate}
\begin{algorithmic}[1]
\Require profile $\profile = (\indexedpaperranking{\setpaperselem{1}}{1}, \ldots, \indexedpaperranking{\setpaperselem{\numreviewers}}{\numreviewers} )$, groups $(\rcx, \pcx), (\rcy, \pcy)$ with $|\pcx|\geq |\pcy|$, aggregation algorithm $\genaggalgo$
\Ensure total ranking of all papers
\State compute $\profile_C$ as the restriction of profile $\profile$ to only papers in $\pcx$, and $\profile_{\bar{C}}$ as the restriction of profile $\profile$ to only papers in $\pcy$
\State  $\rankingx \leftarrow \catname(\profile_C, \genaggalgo)$
\State  $\rankingy \leftarrow \catname(\profile_{\bar{C}}, \genaggalgo)$
\State define $I=\left(\left\lfloor \frac{\numpapers}{|\pcx|} \right\rfloor, \left\lfloor \frac{2\numpapers}{|\pcx|} \right\rfloor,...,\numpapers \right)$
\State {\bf return} total ranking obtained by filling papers in $\pcx$ into positions in $I$ in order given by $\rankingx$, and papers in $\pcy$ into positions in $[\numpapers] \backslash I$ in order given by $\rankingy$\label{step:interleaving}
\\\hrulefill
\Procedure{\catname}{profile $\widetilde{\profile}$, aggregation algorithm $\genaggalgo$}
    \State  build a directed graph $G_{\widetilde{\profile}}$ with the papers in $\widetilde{\profile}$ as its vertices and no edges
    \For {each $i \in [\numreviewers']$}    
        \State denoting $\pi^{(i)} = (\paper{i_1}\succ\ldots \succ \paper{i_{t_i}}$), add a directed edge from $\paper{i_j}$ to $\paper{i_{j+1}}$ in $G_{\widetilde{\profile}}$, $\forall j\in[t_i -1]$\label{step:build_edge}
    \EndFor
    \State  for every ordered pair $(\paper{j_1}, \paper{j_2})\in E_{G_{\widetilde{\profile}}}$, replace multiple edges from $\paper{j_1}$ to $\paper{j_2}$ with a single edge
    \State compute a topological ordering of the strongly connected components (SCCs) in $G_{\widetilde{\profile}}$
    \State for every SCC in $G_{\widetilde{\profile}}$, compute a permutation of the papers in the component using algorithm $\genaggalgo$
    \State \textbf{return} the permutation of all papers that is consistent with the topological ordering of the SCCs and the permutations within the SCCs
\EndProcedure
\end{algorithmic}
\end{algorithm}
At a high level, the papers in each component are aggregated separately using the subroutine \catname. This aggregation in \catname is performed by a topological ordering of all strongly connected components (SCCs) according to the reviews, and then ranking the papers within each set using any arbitrary aggregation algorithm (taken as an input $\genaggalgo$).\footnote{In the case where there are multiple topological orderings, any one of them suffices.} Possible choices for the algorithm $\genaggalgo$ include the modified Borda count~\citep{emerson2013original}, Plackett-Luce aggregation~\citep{hajek2014minimax}, or others~\citep{caragiannis2017optimizing} Moving back to Algorithm \ref{alg:aggregate}, the two rankings returned by \catname respectively for the two components are simply interlaced to obtain a total ranking over all the papers: the slots for $C$ are reserved in set $I$, and $[n]\setminus I$ contain the slots for the remaining papers. 
In our extended version of the paper we also show that the interleaving only causes a small change w.r.t an underlying optimal ranking.

The following theorem now shows that \BiparCATAlgo satisfies group unanimity and is also strategyproof, detailed proof is in Section \ref{sec:proof:thm:gu+ssp}. 
\begin{theorem}
\label{thm:gu+ssp} 
Suppose the vertices of $\conflictgraph$ can be partitioned into two groups $(\rcx, \pcx)$ and $(\rcy, \pcy)$ such that there are no edges in $\conflictgraph$ across the groups and that $\max\big\{ \frac{|\pcx|}{|\rcy|}, \frac{|\pcy|}{|\rcx|}\big\} \leq \frac{\maxpapers}{\minp}$. Then \BiparCATAlgo is group unanimous and strategyproof.
\end{theorem}

\paragraph{Remark.} Our \BiparCATAlgo framework aptly handles the various nuances of real-world conferences peer review, which render other algorithms inapplicable. This includes the aspects that each reviewer can write multiple papers and each paper can have multiple authors, and furthermore that each reviewer may review only a subset of papers. Even under this challenging setting, our algorithm guarantees that no reviewer can influence the ranking of \his own paper via strategic behavior, and it is efficient from a social choice perspective. 

Further, we delve a little deeper into the interleaving step (Step~\ref{step:interleaving}) of the aggregation algorithm. At first glance, this interleaving -- performed independent of the reviewers' reports -- may be a cause of concern. Indeed, assuming there is some ground truth ranking of all papers and even under the assumption that the outputs of the \catname procedure are consistent with this ranking, the worst case scenario is where the interleaving  causes papers to be placed at a positions that are $\Theta(\numpapers)$ away from their respective positions in the true ranking. We show that, however, such a worst case scenario is unlikely to arise, when the  ground truth ranking is independent of the conflict graph. We summarize our findings in the following proposition, with the proof provided later in Section~\ref{sec:proof:thm:misplace}. 

\begin{proposition}
\label{thm:misplace}
Suppose $\conflictgraph$ satisfies the conditions given in Theorem~\ref{thm:gu+ssp} and there exists a constant $\constantnotation\geq 2$ such that $\max\{\frac{|\setpapers|}{|\pcx|}, \frac{|\setpapers|}{|\pcy|}\} \leq \constantnotation$. Assume the ground-truth ranking $\truthfulranking$ is chosen uniformly at random from all permutations in $\allpaperperms{\setpapers}$ independent of $\conflictgraph$, and that the two partial outputs of \catname in Algorithm~\ref{alg:aggregate} respect $\truthfulranking$. Let the output ranking of \BiparCATAlgo be $\estimatedranking$. Then for every $\numpapers \geq 4\constantnotation/\log 2$, for any $\delta\in (0, 1)$, with probability at least $1-\delta$, we have:
\begin{align*}
    \max_{1\leq i\leq \numpapers} |\truthfulranking_i -\estimatedranking_i| \leq 2 \sqrt{\numpapers \constantnotation\cdot \log(2\numpapers/\delta)}.
\end{align*}
\end{proposition}
~\\
Proposition~\ref{thm:misplace} shows that the maximum deviation between the aggregated ranking and the ground truth ranking is $O(\sqrt{n\log(n/\delta)})$ with high probability. Hence when $n$ is large enough, such deviation is negligible when program chairs of conferences need to make accept/reject decisions, where the number of accepted papers usually scales linearly with $n$.

\newcommand{\reviewscore}{o}
\paragraph{Extension to review scores.} Our framework  extends to a score-based setting, wherein each reviewer $\reviewer{\indexreviewer}$ provides their opinion as a score $\reviewscore_{\indexreviewer\indexpaper}$ for every paper $\paper{\indexpaper}\in \setpaperselem{\indexreviewer}$. The assignment algorithm remains the same in this setting; for aggregation, we can use the same procedure with the ranking induced by the review scores. The only difference is that 
in step \ref{step:build_edge} of \catname, we add an edge between every pair of papers $\paper{\indexpaper_1}\rightarrow \paper{\indexpaper_2}$ such that $\reviewscore_{\indexreviewer\indexpaper_1}>\reviewscore_{\indexreviewer\indexpaper_2}$. This makes sure that the graph reflects the opinion of the reviewer and does not impose constraints on papers that are equally rated. In the score-based setting, the aggregation algorithm $\genaggalgo$ is allowed to leverage the review scores for a more granularized ranking (e.g., mean scores). 


\section{Empirical evaluations}

In this section, we perform certain empirical evaluations regarding the feasibility and performance of our \BiparCATAlgo algorithm based on data from the ICLR conference\footnote{The code and data is available at \url{https://github.com/xycforgithub/StrategyProof_Conference_Review}.}. Recall that the \BiparCATAlgo algorithm restricts the assignment of reviewers to papers according to a partition of reviewers and papers into two disconnected groups. By means of these empirical evaluations, we investigate the following questions:
\begin{enumerate}
    \item[Q1.] Is such a partition feasible?
    \item[Q2.] How can one impart more flexibility to the partition (which can allow for better assignments)?
    \item[Q3.] How does the quality of the assignment compare with standard settings without strategyproofness?
    \item[Q4.] One may envisage that reviewers that are more related to the topic of a paper would be more likely to be connected (in the conflict graph) to that paper. Under \BiparCATAlgo, such a reviewer will be barred from being assigned to such a related paper. How much does such a restriction of the assignment between connected reviewers-papers hurt the assignment quality as compared to assignment under a uniform random partition of reviewers and papers?
\end{enumerate}
The most prominent type of conflicts is authorships, and throughout this section we restrict attention to the authorship conflict graph.


\subsection{Analysis of the Conflict Graph on ICLR 2017 submissions (Q1 and Q2)}
\label{sec:analy_iclr_2017}
We address questions Q1 and Q2 using data from the ICLR 2017 conference. In a nutshell:
\begin{enumerate}
    \item[A1.] Yes, partitioning is feasible.
    \item[A2.] We show that removing only a small number of reviewers can result in a dramatic reduction in the size of the largest component in the conflict graph thereby providing great flexibility towards partitioning the papers and reviewers. For instance, removing only $3.5\%$ of all authors from the reviewer pool reduces the size of the largest component (in terms of number of papers)  by 86\%.
\end{enumerate}

We analyze all papers submitted to the ICLR 2017 conference with the given authorship relationship as the conflict graph. ICLR 2017 received 489 submissions by 1417 authors; we believe this dataset is a good representative of a medium-sized modern conference. In the analysis of this dataset, we instantiate the conflict graph as the authorship graph. It is important to note that we consider only the set of authors as the entire reviewer pool (since we do not have access to the actual reviewer identities). Adding reviewers from outside the set of authors would only improve the results since these additional reviewers will have no edges in the authorship conflict graph.

\begin{table}[htb]
\centering
    \caption{Statistics of ICLR 2017 submissions.\label{tab:iclr_17}}
    \begin{tabular}{l|r}\toprule
        Description  & Number  \\\midrule
        Number of submitted papers & 489\\
        Number of distinct authors 
        & 1417\\
        Mean \# papers written per author
        & 1.27\\
        Maximum \# papers written by an author & 14\\\bottomrule
        Number of connected components & 253\\
        \#authors; \#papers in largest connected component & 371; 133\\
        \#authors; \#papers in second largest connected component & 65; 20\\\bottomrule
    \end{tabular}
\end{table}

We first investigate the existence of (moderately sized) components in the conflict graph.  Our analysis shows that the authorship conflict graph is  disconnected, and moreover, has more than 250 components. The largest connected component (CC) contains 133 (that is, about $27\%$) of all papers, and the second largest CC is much smaller. 
We tabulate the results from our analysis in Table~\ref{tab:iclr_17}. These statistics indeed verify our assumption in Theorem~\ref{thm:gu+ssp} that the conflict graph is disconnected and can be divided into two disconnected parts of similar size. 

The partitioning method has previously been considered for the problem of peer grading~\citep{kahng2017ranking}. The peer grading setting is  homogeneous in that each reviewer (student) goes through the same course and hence any paper (homework) can be assigned to any reviewer. In peer review, however, different reviewers typically have different areas of expertise and hence their abilities to review any paper varies by the subject area of the paper. In order to accommodate this diversity in area of expertise in peer review, one must have a greater flexibility in terms of assigning papers to reviewers. In our analysis in Table~\ref{tab:iclr_17} we saw that the largest connected component comprises 372 authors and 133 papers. It is reasonable to expect that a large number of reviewers with expertise required to review these 133 papers may fall in the same connected component, meaning that a na\"ive application of \BiparCATAlgo to this data would assign these 133 papers to reviewers who may have a lower expertise for these papers. This is indeed a concern, and in what follows, we discuss a simple yet effective way to ameliorate this problem.

A simple yet (as we demonstrate below) effective idea is to remove some authors from the reviewer pool. Empirically using the ICLR 2017 data, we show that by removing only a small number of authors from the reviewer pool, we can make the conflict graph considerably more sparse, thereby allowing for a significantly more flexible application of our algorithm \BiparCATAlgo (or more generally, any partition-based algorithm). We use the simple heuristic of removing the authors with the maximum degree in the (authorship) conflict graph. We then study the resulting conflict graph (containing all submitted papers but only the remaining reviewers) in terms of the numbers and sizes of the connected components. We present the results in Table~\ref{tab:iclr_remove}. We see that on removing only a small fraction of authors --- 50 authors which is only about $3.5\%$ of all authors --- the number of papers in the largest connected component reduces by 86\% to just 18. Likewise, the number of authors in the largest connected component reduces to 55 from 371 originally. 

\begin{table}[htb]
\caption{Statistics of the conflict graph on removing a small number ($< 7\%$) of authors from the reviewer pool comprising the 1417 authors.\label{tab:iclr_remove}}
    \centering
    \begin{tabular}{l|c|c|c|c|c|c|c}\toprule
     & \multicolumn{7}{c}{\#Authors removed from reviewer pool}\\
     & 0 & 5 & 10 & 15 & 20 & 50 & 100\\\midrule
     Number of Components & 253 & 268 & 278 & 292 & 302 & 334 & 389\\
     Number of Authors in Largest CC  & 371 & 313 & 304 & 228 & 205 & 55 & 28\\
     Number of Papers in Largest CC   & 133 & 114 & 110 & 82 & 74 & 18 & 8\\\bottomrule
    \end{tabular}
\end{table} 

\subsection{Analysis of the Partition Algorithm on ICLR 2018 submissions (Q3 and Q4)}
\label{sec:analy_iclr_2018}
In the previous section, we empirically verified that we can partition the reviewers and papers into two disconnected groups.
A natural question that arises is how such a partition affects to the overall reviewer-paper matching process, i.e., will the partition cause a great loss in quality of the assignment algorithm used in practice? We empirically investigate this question by performing experiments using data from the ICLR 2018 conference (where we also use ICLR 2017 as a reference point later). In a nutshell, using the popular ``mean similarity score'' as a measure of the quality of the assignment (detailed below), we see that:
\begin{itemize}
    \item[A3.] In comparison to when there is no strategyproofing, the quality of the assignment reduces by $11\%$ when it is made strategyproof using the \partition algorithm.
    \item[A4.] The utility under \partition is only marginally lower than when the partition is done uniformly at random. Thus \partition is roughly equivalent to shrinking the conference size (randomly) to a half.
\end{itemize}

\newcommand{\simscore}{s}
We now describe the experiment in more detail. We follow the assignment framework used popularly in practice, which comprises of two phases. The first phase computes a similarity score for every (reviewer, paper) pair.
A higher similarity score is interpreted as a higher envisaged quality of review. As in the case of ICLR 2017 above, we set the collection of all authors as the reviewer pool since the actual identities of the collection of reviewers are not available. We then compute a similarity between every reviewer-paper pair based on the text of the paper and the contents of the reviewer's published papers. We refer the reader to Appendix~\ref{sec:appexp} for details of this construction.

\begin{figure}[htb]
    \centering
    \begin{subfigure}[t]{0.48\linewidth}
        \includegraphics[width=\linewidth]{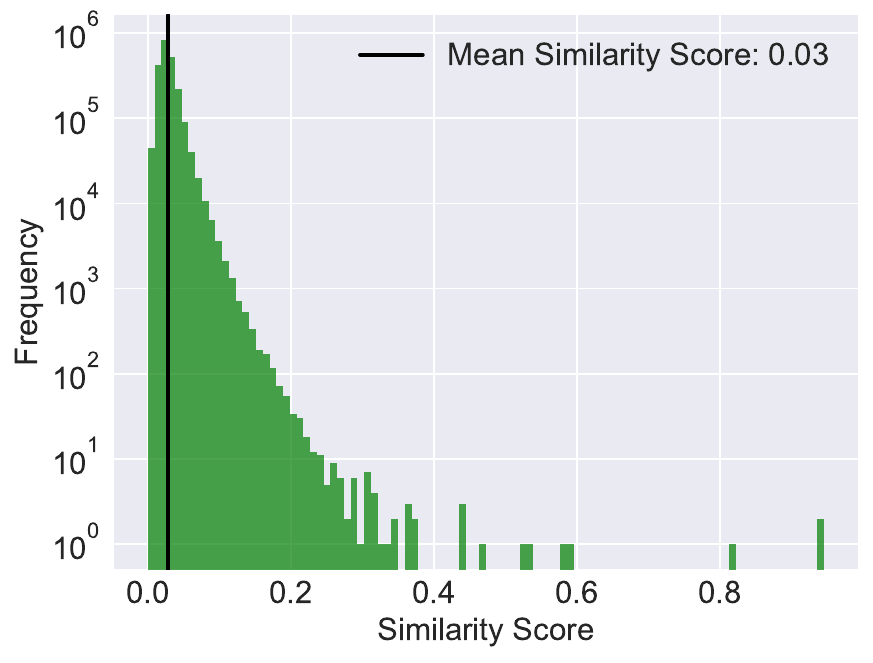}
        \caption{Scores between reviewers and papers.}
        \label{fig:score}
    \end{subfigure}
    ~
    \begin{subfigure}[t]{0.48\linewidth}
        \includegraphics[width=\linewidth]{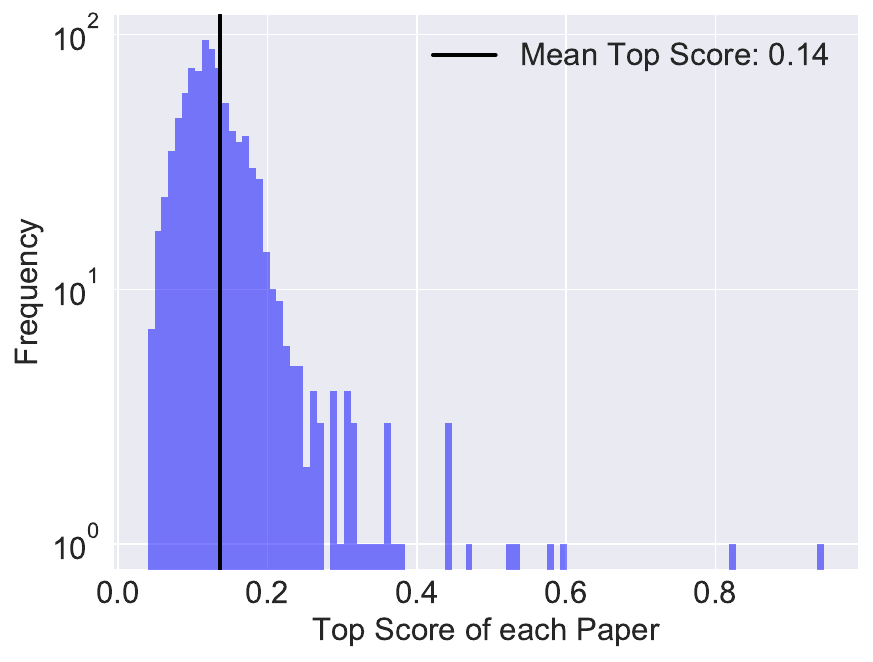}
        \caption{Top score of each paper.}
        \label{fig:top}
    \end{subfigure}
    \caption{The left histogram is the similarity scores between reviewers and papers of the ICLR 2018 dataset. The right histogram plots the top similarity score of each paper in the ICLR 2018 dataset. In both plots, the vertical black line shows the mean of the distribution.}
\end{figure}

Here are some basic statistics about the computed similarity matrix. In Figure~\ref{fig:score} we plot the histogram of the computed similarity scores between 911 papers and 2435 reviewers. The mean of the similarity scores across all reviewer-paper pairs is approximately $0.03$. This skewed distribution of similarity scores is consistent with our intuition: for each paper, there is only a handful of reviewers who have the aligned background and expertise. In Figure~\ref{fig:top}, we show the histogram of the top similarity score computed for each paper (excluding reviewers that are also authors of the corresponding paper). We see that the mean (across all papers) of these top scores is approximately 0.14, which is significantly higher than that of 0.03 among all similarity scores.  

The second phase of the assignment procedure uses the similarity scores to assign reviewers to papers. The most widely used assignment method used in practice is the Toronto Paper Matching System or TPMS~\citep{charlin13tpms}. This assignment method maximizes the mean similarity score across all assigned reviewer-paper pairs. 

In what follows, we evaluate three different methods of assigning reviewers to papers in terms of the resulting mean similarity score across all assigned reviewer-paper pairs:
\begin{itemize}
    \item Using TPMS assignment without any partitioning.
    \item The \BiparCATAlgo (i.e., partitioning reviewers and papers into two sets that are disconnected in the authorship conflict graph) with TPMS as the assignment algorithm $\genassalgo$.
    \item Partitioning the reviewers and papers into two equal groups uniformly at random, and using TPMS with the restriction of assigning each reviewers to papers from the other group.
\end{itemize}
We use the values $\mu=6, \lambda=3$ which are typical of conferences today. 
We provide specific implementation details in Appendix~\ref{sec:appexp}.


Basic statistics and experiment results are shown in Table \ref{tab:iclr_18}. ICLR has grown dramatically from 2017 to 2018, with the number of papers rising from 489 to 911, and the corresponding numbers of authors and components also almost double. The largest component now has 757 authors and 274 components, twice the size of 2017. On the other hand, the second largest component is smaller than that of 2017, which we speculate is because the machine learning community has grown into more refined subfields, creating more smaller clusters. Nevertheless, we are still able to divide the authors and papers into two clusters of approximately equal size using \partition.

\begin{table}[htb]
\centering
    \caption{Result of experiments on ICLR 2018 data.\label{tab:iclr_18}}
    \begin{tabular}{lr}\toprule
        Description  & Number  \\\midrule
        Number of submitted papers & 911\\
        Number of distinct authors & 2428\\
        Number of connected components & 465\\
        \#authors, \#papers in largest connected component & 757, 274\\
        \#authors, \#papers in second largest connected component & 30, 11\\\midrule
        Mean similarity score without partitioning & 0.0880        \\
        Mean similarity score with random partition (20 runs) & $0.0779\pm0.0001$\\
        Mean similarity score with \partition & 0.0782\\
        \bottomrule
    \end{tabular}
\end{table}

The mean of similarity scores using TPMS score assignment (see Appendix \ref{app:assignment_tpms}) without partitioning is 0.0880 and that with partition is 0.0782, which represents a $11.4\%$ decrease from partitioning. On the other hand, if we randomly partition the reviewers and papers to two sets of equal sizes, the mean (and standard deviation) of the similarity is $0.0779$ (and $0.0001$ respectively) from 20 runs.
The result from \partition is similar to the result obtained from random partition. So in terms of author-paper similarities, \partition does no additional harm than shrinking the conference size (randomly) to a half. For ICLR 2018, this just corresponds to the size of ICLR 2017.

\section{Negative Theoretical Results}
\label{sec:neg}

The positive results in the previous section focus on group unanimity, which is weaker than the conventional notion of unanimity (the conventional notion is also known as pairwise unanimity). Moreover, the algorithm had a disconnected review graph whereas the review graphs of conferences today are typically connected \citep{shah2017design}. It is thus natural to wonder about the extent to which these results can be strengthened: Can a peer-review system with a connected reviewer graph satisfy these properties? Can a strategyproof peer-review system be pairwise unanimous? In this section we present some negative results toward these questions, thereby highlighting the critical impediments towards (much) stronger results. 

Before stating our results, we introduce another notion of strategyproofness, which is significantly weaker than the notion of \SPl (Definition~\ref{def:s-strategyproof}), and is hence termed as \WSPl. As compared to \SPl which is defined with respect to a given conflict graph, \WSPl only requires the existence of a conflict graph (with non-zero reviewer-degrees) for which the review process is strategyproof.
\begin{definition}[Weak Strategyproofness, \WSP]
\label{def:w-strategyproof}
A review process $(\pairopt)$ is called \emph{\WSPlv}, if
for every reviewer $\reviewer{\indexreviewer}$, there exists some paper $\paper{\indexpaper}\in \setpapers$ such that for every pair of distinct profiles (under assignment $\revgraph$) $\profile = (\indexedpaperrankshort{1},\ldots,\indexedpaperrankshort{i-1},\indexedpaperrankshort{i},\indexedpaperrankshort{i+1},\ldots,\indexedpaperrankshort{\numreviewers})$ and $\profile' = (\indexedpaperrankshort{1},\ldots,\indexedpaperrankshort{i-1},{\indexedpaperrankshort{i}}',\indexedpaperrankshort{i+1},\ldots,\indexedpaperrankshort{\numreviewers})$, it is guaranteed that $\aggfunctionnotation_{\indexpaper}(\profile) = \aggfunctionnotation_{\indexpaper}(\profile')$.
\end{definition} 

In other words, \WSPl requires that for each reviewer there is at least one paper (not necessarily shares conflicts this reviewer) whose ranking cannot be influenced by the reviewer. As the name suggests, \SSPl is strictly stronger than \WSPl, when each reviewer has at least one paper of conflict.

We define the notion of weak strategyproofness mainly for theoretical purposes to establish negative results, since \WSP is too weak to be practical. However, even this extremely weak requirement is impossible to satisfy in situations of practical interest. 

\begin{table}[H]
    \centering
    \caption{Summary of our negative results (first three rows of the table), and a comparison to our positve result (fourth row).}
    \label{tab:all_results}
    \begin{tabular}{|c|c|c|c|c|}
    \hline
        Unanimity & Strategyproof & Requirement on $\revgraph$ & Possible? & Reference\\
        \hline
        Pairwise & None & Mild (see Corollary \ref{corol:neg_pu}) & No & Theorem \ref{thm:neg_pu} \\
        Group & Weak & Mild (Connected $\revgraph$) & Conjecture: No & Proposition \ref{prop:exampleguwsp} \\
        Pairwise & Weak & Complete $\revgraph$ & No & Theorem \ref{thm:neg_pu_total}\\
        \hdashline
         Group & Yes & None & Yes & Theorem \ref{thm:gu+ssp}\\
         \hline
    \end{tabular}
\end{table}

We summarize our results in Table \ref{tab:all_results}. Recall that we show the property of group unanimity and strategyproof for \BiparCATAlgo; as the first direction of possible extension, we show in Theorem \ref{thm:neg_pu} that the slightly stronger notion of \PUln is impossible to satisfy under mild assumptions, even \emph{without} strategyproof constraints. Then in Section \ref{sec:neg_gu_wsp} we explore the second direction of extension, by requiring a connected $\revgraph$; we give conjectures and insights that group unanimity and weak strategyproofness is impossible under this setting. At last in Theorem \ref{thm:neg_pu_total} we revert to the traditional setting of social choice, where every reviewer gives a total ranking of the set of all papers $\setpapers$; we show that in this setting it is impossible for any review process to be pairwise unanimous and weakly strategyproof.

\subsection{Impossibility of Pairwise Unanimity \label{sec:neg_pu}}
We show in this section that \PUln is too strong to satisfy under mild assumptions. These assumptions are mild in the sense that a violation of the assumptions leads to severely limited and somewhat impractical choices of $\revgraph$. 
\newcommand{\revrelgraph}{\mathcal{H}}

In order to precisely state our result, we first introduce the notion of a \emph{review-relation graph} $\revrelgraph$. Given a paper-review assignment $\{\setpaperselem{\indexreviewer}\}_{\indexreviewer=1}^{\numreviewers}$, the review-relation graph $\revrelgraph$ is an undirected graph with $[\numpapers]$ as its vertices and where any two papers $\paper{\indexpaper_1}$ and $\paper{\indexpaper_2}$ are connected iff there exists at least one reviewer who reviews both the papers. With this preliminary in place, we are now ready to state the main result of this section:

\begin{theorem}
\label{thm:neg_pu}
If $\revrelgraph$ has a cycle of length 3 or more and there is no single reviewer reviews all the papers in the cycle, then there is no review process $(\pairopt)$ that is \PUl.
\end{theorem}
\newcommand{\workload}{\mu}
The proof of Theorem~\ref{thm:neg_pu} is similar to a Condorcet cycle proof, and the details are in Section~\ref{sec:proof:thm:neg_pu}. In the corollary below we give some direct implications of the condition in Theorem~\ref{thm:neg_pu} when $|\setpaperselem{1}|=\cdots =|\setpaperselem{\numreviewers}|=\workload$, that is, when every reviewer ranks a same number of papers.
\begin{corollary}
\label{corol:neg_pu}
Suppose $|\setpaperselem{1}|=\cdots =|\setpaperselem{\numreviewers}|=\workload\geq 2$. If $(\pairopt)$ is \PUl, the following conditions hold:
\begin{enumerate}[(i)]
    \item \label{point0:neg_pu} $\revrelgraph$ does not contain any cycles of length $\workload+1$ or more.
    \item \label{point1:neg_pu} The set of papers reviewed by any pair of reviewers  $\reviewer{\indexreviewer_1}$ and $\reviewer{\indexreviewer_2}$ must satisfy the condition $|\setpaperselem{\indexreviewer_1}\cap \setpaperselem{\indexreviewer_2}|\in \{0,1,\workload\}$. In words, if a pair of reviewers review more than one common papers, they must review exactly the same set.
    \item \label{point2:neg_pu} The number of distinct sets in $\setpaperselem{i}, \ldots, \setpaperselem{\numreviewers}$ is at most $\frac{\numpapers-1}{\workload-1}$.
\end{enumerate}
\end{corollary}

\paragraph{Remarks.} In modern conferences~\citep{shah2017design}, each reviewer usually reviews around 3 to 6 papers. If we make the review process pairwise unanimous, by~Corollary \ref{corol:neg_pu}~\ref{point2:neg_pu} the number of distinct review sets is much smaller than the number of reviewers; this severely limits the design of review sets, since many reviewers would be necessitated to review identical sets of papers. ~Corollary \ref{corol:neg_pu}~\ref{point1:neg_pu} is a related, strong requirement, since the specialization of reviewers might not allow for such limiting of the intersection of review sets. For instance, there are a large number of pairs of reviewers who review more than one common paper but none with exactly the same set of papers~\citep{shah2017design}. 

In summary, Theorem~\ref{thm:neg_pu} and Corollary~\ref{corol:neg_pu} show that it is difficult to satisfy \PUln, even without considering strategyproofness. This justifies our choice of group unanimity in the positive results.


\subsection{Group Unanimity and Strategyproof for a Connected Review Graph\label{sec:neg_gu_wsp}}

Having shown that \PUln is too strong a requirement to satisfy, we now consider another direction for extension -- conditions on the review graph $\revgraph$. A natural question follows: Under what condition on the review graph $\revgraph$ are  both \GUln and \SPl possible? Although we will leave the question of finding the exact condition open, we conjecture that if we require $\revgraph$ to be connected, then group unanimity and strategyproofness cannot be simultaneously satisfied. To show our insights, we analyze an extremely simplified review setting. 
\newcommand{\tempsetpapers}[1]{P_{#1}}

\begin{proposition}
    Consider any $\numpapers \geq 4$ and suppose $\setpapers=\tempsetpapers{1}\cup \tempsetpapers{2} \cup \tempsetpapers{3} \cup \tempsetpapers{4}$, where $\tempsetpapers{1},\tempsetpapers{2},\tempsetpapers{3},\tempsetpapers{4}$ are disjoint nonempty sets of papers. Consider a review graph $\revgraph$ with $\numreviewers=3$ reviewers, where reviewer $\reviewer{1}$ reviews $\{\tempsetpapers{1},\tempsetpapers{2}\}$, $\reviewer{2}$ reviews  $\{\tempsetpapers{2},\tempsetpapers{3}\}$, and $\reviewer{3}$ reviews $\{\tempsetpapers{3},\tempsetpapers{4}\}$. Then there is no aggregation function $\aggfunctionnotation$ that is both weakly strategyproof and group unanimous.
    \label{prop:exampleguwsp}
\end{proposition}

Proposition \ref{prop:exampleguwsp} thus shows that for the simple review graph considered in the statement, group unanimity and \WSPl cannot hold at the same time. Proof details can be found in Section~\ref{sec:proof:prop:exampleguwsp}. 

We conjecture that such a negative result may hold for more general connected review graphs, and such a negative result may be proved by identifying a component of the general review graph that meets the condition of Proposition~\ref{prop:exampleguwsp}. This shows that our design process of the review graph in Section~\ref{sec:pos} is quite essential for ensuring those important properties.

\subsection{Pairwise Unanimity and Strategyproof under Total Ranking\label{sec:neg_pu_totalrank}}

Throughout the paper so far,  motivated by the application of conference peer review, we considered a setting where every reviewer reviews a (small) subset of the papers. In contrast, a bulk of the classical literature in social choice theory considers a setting where each reviewer ranks \emph{all} candidates or papers~\citep{arrow1950difficulty,satterthwaite1975strategy}. Given this long line of literature, intellectual curiosity drives us to study the case of all reviewers reviewing all papers for our conference peer-review setting. 

We now consider our notion of \PUl and weakly \SPlv in this section under this total-ranking setting, where $\setpaperselem{1}=\cdots =\setpaperselem{\numreviewers}=\setpapers$. In this case, the review graph $\revgraph$ is always a complete bipartite graph, and it only remains to design the aggregation function $\aggfunctionnotation$. Although total rankings might not be practical for large-scale conferences, it is still helpful for smaller-sized conferences and workshops.

Under this total ranking setting, we prove a negative result showing that pairwise unanimity and strategyproofness cannot be satisfied together, and furthermore, even the notion of weak strategyproofness (together with \PU) is impossible to achieve. 
\begin{theorem}
\label{thm:neg_pu_total}
Suppose $\numpapers\geq 2$. If $\setpaperselem{1}=\cdots =\setpaperselem{\numreviewers}=\setpapers$, 
then there is no aggregation function $\aggfunctionnotation$ that is both no aggregation function $\aggfunctionnotation$ that is both and pairwise unanimous.
\end{theorem}
Note that the conditions required for Theorem~\ref{thm:neg_pu} are not met in the total ranking case.
To prove Theorem \ref{thm:neg_pu_total}(details in Section~\ref{sec:prooflemdiagonal}), we use Cantor's diagonalization argument to generate a contradiction by assuming there exists $\aggfunctionnotation$ that is both \PUs and \WSPs.

It is interesting to note that pairwise unanimity can be easily satisfied in this setting of total rankings, by using a simple aggregation scheme such as the Borda count. However, Theorem~\ref{thm:neg_pu_total} shows that surprisingly, even under the extremely mild notion of strategyproofness given by \WSP, it is impossible to achieve pairwise unanimity and strategyproofness simultaneously.


\section{\label{sec:proof}Proofs}

In this section, we provide the detailed proofs of all the results from previous sections. 

\subsection{Proof of Proposition \ref{lemma:unanimity} \label{sec:proof:lemma:unanimity}}
Suppose $(\pairopt)$ is \PUs, and $\setpapers' \subset \setpapers$ satisfies that every reviewer ranks the papers \he reviewed from $\setpapers'$ higher than those \he reviewed from $\setpapers\setminus \setpapers'$. Now for every $\paper{x}\in \setpapers'$ and $\paper{y}\in \setpapers\setminus \setpapers'$ and reviewer $\reviewer{\indexreviewer}$ such that $\reviewer{\indexreviewer}$ reviews both $\paper{x}$ and $\paper{y}$, $\reviewer{\indexreviewer}$ must rank $\paper{x}\succ \paper{y}$ since otherwise the assumption of $\setpapers'$ is violated. Since $\aggfunctionnotation$ is \PUs, we know that $\aggfunction{\profile}$ must respect $\paper{x}\succ \paper{y}$ as well. This argument holds for every $\paper{x}\in \setpapers'$ and $\paper{y}\in \setpapers\setminus \setpapers'$ that have been reviewed by at least one reviewer, and hence $(\pairopt)$ is also \GUs.

\subsection{Proof of Theorem~\ref{thm:gu+ssp} \label{sec:proof:thm:gu+ssp}}

We assume that the condition on the partitioning of the conflict graph, as stated in the statement of this theorem, is met. We begin with a lemma which shows that for any aggregation algorithm $\genaggalgo$, \catname is group unanimous. 

\begin{lemma}
\label{lemma:cat}
For any assignment and aggregation algorithms $\genassalgo$ and $\genaggalgo$, the aggregation procedure \catname is \GUl. 
\end{lemma}
We prove this lemma in Section~\ref{sec:prooflemmacat}. Under the assumptions on $\maxpapers$, $\minp$ and sizes of $\rcx,\rcy, \pcx, \pcy$, it is easy to verify that there is a paper allocation satisfies $|\setpapers_i|\leq \maxpapers, \forall~i \in [\numreviewers]$ and each paper gets at least $\minp$ reviews. The strategyproofness of \BiparCATAlgo follows from the standard ideas in the past literature on partitioning-based methods~\citep{alon2011sum}: Algorithm~\ref{alg:assign} guarantees that reviewers in $\rcx$ do not review papers in $\pcy$, and reviewers in $\rcy$ do not review papers in $\pcx$. Hence the fact that \BiparCATAlgo is strategyproof trivially follows from the assignment procedure where each reviewer does not review the papers that are in conflict with her, as specified by the conflict graph $\conflictgraph$. Given that all the other reviews are fixed, the ranking of the papers in conflict with her will only be determined by the other group of reviewers and so fixed no matter how \he changes \his own ranking. On the other hand, from Lemma~\ref{lemma:cat}, since \catname is group unanimous, we know that $\rankingx$ and $\rankingy$ respect group unanimity w.r.t. $\profile_{C}$ and $\profile_{\bar{C}}$, respectively. Since $\profile = (\profile_{C}, \profile_{\bar{C}})$, it follows that $\rankingx$ and $\rankingy$ also respect group unanimity w.r.t. $\profile$. Finally, note that there is no reviewer who has reviewed both papers from $\pcx$ and $\pcy$, the interlacing step preserves the group unanimity, which completes our proof.


\subsubsection{Proof of Lemma \ref{lemma:cat}}
\label{sec:prooflemmacat}
Let $\aggfunctionnotation(\widetilde{\profile}) \defn \catname(\widetilde{\profile}, \genaggalgo)$, where $\widetilde{\profile}$ is a preference profile. Define $\pi = \aggfunctionnotation(\widetilde{\profile})$. Let $k$ denote the number of SCCs in $G_{\widetilde{\profile}}$. Construct a directed graph  $\widetilde{G}_{\widetilde{\profile}}$ such that each of its vertices represents a SCC in $G_{\widetilde{\profile}}$, and there is an edge from one vertex to another in $\widetilde{G}_{\widetilde{\profile}}$ iff there exists an edge going from one SCC to the other in the original graph $G_{\widetilde{\profile}}$. Let $\tilde{v}_1, \ldots, \tilde{v}_k$ be a topological ordering of the vertices in $\widetilde{G}_{\widetilde{\profile}}$. Since $\tilde{v}_1, \ldots, \tilde{v}_k$ is a topological ordering, then edges can only go from $\tilde{v}_{j_1}$ to $\tilde{v}_{j_2}$ where $j_1 < j_2$. Now consider any cut $(\setpapers_X, \setpapers_Y)$ in $G_{\widetilde{\profile}}$ that satisfies the requirement of group unanimity, i.e., all edges in the cut direct from $\setpapers_X$ to $\setpapers_Y$. Then there is no pair of papers $\genpaper_x\in \setpapers_X$ and $\genpaper_y\in \setpapers_Y$ such that $\genpaper_x$ and $\genpaper_y$ are in the same connected component, otherwise there will be both paths from $\genpaper_x$ to $\genpaper_y$ and $\genpaper_y$ to $\genpaper_x$, contradicting that $(\setpapers_X, \setpapers_Y)$ forms a cut where all the edges go in one direction. This shows that $\setpapers_X$ and $\setpapers_Y$ also form a partition of all the vertices $\tilde{v}_1, \ldots, \tilde{v}_k$. Now consider any edge $(\genpaper_x,\genpaper_y)$ from $\setpapers_X$ to $\setpapers_Y$. Suppose $\genpaper_x$ is in component $\tilde{v}_{j_x}$ and $\genpaper_y$ in component $\tilde{v}_{j_y}$. We have $j_x\ne j_y$, since $\setpapers_X$ and $\setpapers_Y$ forms a partition of all SCCs; also it cannot happen that $j_x>j_y$, otherwise $\tilde{v}_1,\ldots,\tilde{v}_k$ is not a topological ordering returned by $\aggfunctionnotation$. So it must be $j_x<j_y$, and the edge $(\genpaper_x,\genpaper_y)$ is respected in the final ordering.

\subsection{\label{sec:proof:thm:misplace}Proof of Proposition \ref{thm:misplace}}
We would first need a lemma for the location of papers:
\newcommand{\sfloor}{g}
\begin{lemma}
\label{lemma:techpos}
Let 
$I_1=\left\{\left\lfloor \frac{\numpapers}{|\pcx|} \right\rfloor, \left\lfloor \frac{2\numpapers}{|\pcx|} \right\rfloor,...,\numpapers \right\}, I_2=\left\{\sfloor\left( \frac{\numpapers}{|\pcy|} \right), \sfloor\left( \frac{2\numpapers}{|\pcy|} \right),...,\numpapers-1\right\}$, where $\sfloor(x)=\lceil x\rceil -1$ is the largest integer that is \emph{strictly} smaller than $x$. Then $I_1\cap I_2=\emptyset$ and $I_1\cup I_2=[\numpapers]$.

\end{lemma}
We prove this lemma in Section~\ref{sec:prooflemtechpos}.

Consider any paper $\paper{i}$, and suppose its position in $\truthfulranking$ is $\paperpos$. Define $n_1 = |\setpapers_C|$ and $n_2 = |\setpapers_{\bar{C}}|$. Without loss of generality assume $n_1 \geq n_2$ (the other case is symmetric) and let $\numpapers \geq 4\constantnotation / \log 2$. We discuss the following two cases depending on whether $\paper{i}\in \setpapers_C$ or $\paper{i}\in \setpapers_{\bar{C}}$. 

\textbf{Case I}: If $\paper{i}\in\setpapers_C$. Let $\numberBeforePaperi$ be the number of papers in $\paperC$ ranked strictly higher (better) than $\paperpos$ according to $\truthfulranking$. Since the permutation $\truthfulranking$ is uniformly random, conditioned on this value of $\paperpos$, the other papers' positions in the true ranking are uniformly at random in positions $[\numpapers]\setminus \{\paperpos\}$. Now for any paper $\paper{j}, j\neq i$, let $\randomvar_j$ be an indicator random variable set as 1 if position of $\paper{j}$ is higher than $\paperpos$ in $\truthfulranking$, and 0 otherwise. So $\numberBeforePaperi=\sum_{\paper{j}\in \paperC\backslash\{\paper{i}\}} \randomvar_j$, and $\Pr(\randomvar_j=1)=\frac{\paperpos-1}{\numpapers-1}$ when $j \neq i$.  Then using Hoeffding's inequality without replacement, we have
\[\Pr\left(\left|\frac{\numberBeforePaperi}{n_1 -1}- \frac{\paperpos-1}{\numpapers-1}\right|\geq \varepsilon \right)\leq 2\exp(-2(n_1 - 1)\varepsilon^2)\leq 2\exp(-n_1\varepsilon^2) \]
for any $\varepsilon>0$. The last inequality is due to $n_1 \geq 2$, which holds because $\numpapers/n_1 \leq \constantnotation$ with a constant $\constantnotation$. Now setting $\varepsilon=\sqrt{\frac{\log (2/\delta)}{n_1}} $ we have the bound
\[\Pr \left( \left| \frac{\numberBeforePaperi}{n_1 -1}- \frac{\paperpos-1}{\numpapers-1}\right| \leq \sqrt{\frac{\log (2/\delta)}{n_1}} \right) \geq 1 - \delta. \]
Now note that by Algorithm~\ref{alg:aggregate}, the position of paper $\paper{i}$ in the ranking $\estimatedranking$ is $\estimatedranking_i = \Big\lfloor (\numberBeforePaperi+1)\cdot \frac{\numpapers}{n_1} \Big\rfloor$. Use this relationship to substitute $k$ in the above inequality, and notice that by assumption $\max\{\numpapers / n_1, \numpapers / n_2\} \leq \constantnotation$, we have
\begin{align*}
    (\numberBeforePaperi+1)\cdot \frac{\numpapers}{n_1}&\leq \left((n_1-1)\left(\varepsilon+\frac{\paperpos-1}{\numpapers-1}\right)+1\right)\frac{\numpapers}{n_1}\\
    &=\frac{n_1-1}{n_1}\cdot \frac{\numpapers}{\numpapers-1}(\paperpos-1)+\frac{n_1-1}{n_1}\cdot \numpapers\varepsilon+\frac{\numpapers}{n_1}\\
    &\leq \frac{\numpapers}{n_1}+\numpapers\varepsilon+\paperpos-1.
\end{align*}
On the other hand,
\begin{align*}
    (\numberBeforePaperi+1)\cdot \frac{\numpapers}{n_1}&\geq \left((n_1-1)\left(\frac{\paperpos-1}{\numpapers-1}-\varepsilon\right)+1\right)\frac{\numpapers}{n_1}\\
    &=\frac{n_1-1}{n_1}\cdot \frac{\numpapers}{\numpapers-1}(\paperpos-1)-\frac{n_1-1}{n_1}\cdot \numpapers\varepsilon+\frac{\numpapers}{n_1}\\
    &\geq \frac{\numpapers}{n_1}-\numpapers\varepsilon+\frac{n_1-1}{n_1}\cdot \frac{\numpapers}{\numpapers-1}(\paperpos-1).
\end{align*}
So
\begin{align}
  (\numberBeforePaperi+1)\cdot \frac{\numpapers}{n_1}-\paperpos&\geq \frac{\numpapers}{n_1}-\numpapers\varepsilon+\frac{n_1-1}{n_1}\cdot \frac{\numpapers}{\numpapers-1}(\paperpos-1)-\paperpos  \label{eqn:geq1}\\
  &\geq \frac{\numpapers}{n_1}-\numpapers\varepsilon+\frac{n_1-1}{n_1}\cdot \frac{\numpapers}{\numpapers-1}(\numpapers-1)-\numpapers  \label{eqn:geq2}\\
  &\geq -\numpapers\varepsilon.\nonumber
  \end{align}
  Here (\ref{eqn:geq2}) is because $\frac{n_1-1}{n_1}\cdot \frac{\numpapers}{\numpapers-1}< 1$, and thus RHS of (\ref{eqn:geq1}) is minimized (as a function of $\paperpos$) when $\paperpos=\numpapers$. Combining the two inequalities above we have
\[|\estimatedranking_i -\paperpos|\leq \numpapers\varepsilon+\frac{\numpapers}{n_1} = \numpapers\sqrt{\frac{\log(2/\delta)}{n_1}}+\frac{\numpapers}{n_1}\leq 2\sqrt{\numpapers \constantnotation\cdot\log(2/\delta)}, \]
where the last inequality is by the assumption that $\numpapers$ is large enough so that $2\constantnotation \leq \sqrt{\numpapers\constantnotation\cdot \log(2/\delta)}$.

\textbf{Case II}: If $\paper{i}\in \setpapers_{\bar{C}}$. Again, let $\numberBeforePaperi$ be the number of papers in $\setpapers_{\bar{C}}$ ranked strictly higher (better) than $\paperpos$ according to $\truthfulranking$. As the analysis in Case I, similarly, we have $\numberBeforePaperi=\sum_{\paper{j}\in \setpapers_{\bar{C}}\backslash\{\paper{i}\}} \randomvar_j$, and $\Pr(\randomvar_j=1)=\frac{\paperpos-1}{\numpapers-1}$. With the same analysis using Hoeffding's inequality without replacement, with probability at least $1-\delta$ we have
\[\left|\frac{\numberBeforePaperi}{n_2 -1}- \frac{\paperpos-1}{\numpapers-1}\right| \leq \sqrt{\frac{\log (2/\delta)}{n_2}}. \]
Now using Lemma \ref{lemma:techpos}, the position of paper $\paper{i}$ in $\estimatedranking$ in this case is $\estimatedranking_i = \sfloor\left((\numberBeforePaperi+1)\cdot \frac{\numpapers}{n_2}\right)$. Using exactly the same analysis as Case I we have
\[-\numpapers\varepsilon\leq (\numberBeforePaperi+1)\cdot \frac{\numpapers}{n_2}-\paperpos\leq \frac{\numpapers}{n_2}+\numpapers\varepsilon-1, \] 
and thus
\[|\estimatedranking_i -\paperpos|\leq \numpapers\varepsilon+\frac{\numpapers}{n_2} = \numpapers\sqrt{\frac{\log(2/\delta)}{n_2}}+\frac{\numpapers}{n_2}\leq 2\sqrt{\numpapers \constantnotation\cdot\log(2/\delta)}. \]

Combine both Case I and Case II, and notice that $\truthfulranking_i$ is uniformly distributed in $[\numpapers]$. Using a union bound over $i=1,2,...,\numpapers$, with probability $1-\delta$ we have:
\[\max_{1\leq i\leq \numpapers}|\estimatedranking_i -\truthfulranking_i|\leq 2 \sqrt{\numpapers \constantnotation\cdot\log(2\numpapers/\delta)}.\]

\subsubsection{Proof of Lemma \ref{lemma:techpos}}
\label{sec:prooflemtechpos}
We show that for every slot $\indexrankk$ that there is no $\indexrank$ such that $\newfloorfunc{\indexrank\cdot \frac{\numpapers}{n_1}}=\indexrankk$, there exists one slot $\indexrank'$ for $\pcy$ such that $\sfloor\left( \indexrank'\cdot\frac{\numpapers}{n_2} \right)=\indexrankk$, i.e., all slots that are left empty by $\pcx$ are taken by slots of $\pcy$. Since that the two kinds of slots have a total number of $n$, we show that there are no overlap between the two kinds of slots, thus proving the lemma.

Let $t=\numpapers/n_1$. Suppose if there is no $\indexrank$ such that $\newfloorfunc{\indexrank\cdot \frac{\numpapers}{n_1}}=\indexrankk$, then  there must exist some $\hat{\indexrank}$ such that 
\begin{align}
    \indexrankk+1\leq \hat{\indexrank} t<\indexrankk+t.\label{eqn:rel}
\end{align}
This is because there must be a multiple of $t$ in the range $[\indexrankk,\indexrankk+t)$, but our assumption makes that there is no such multiply in $[\indexrankk,\indexrankk+1)$. 

Now let $u=\numpapers/n_2$. By $n_1+n_2=\numpapers$ we have $1/u+1/t=1$; substituting $t=u/(u-1)$ in (\ref{eqn:rel}) we have
\[\indexrankk<(\indexrankk-\hat{\indexrank}+1)u\leq \indexrankk+1.\]
Thus there exists $p'=\sfloor((\indexrankk-\hat{\indexrank}+1)u)\in \pcy$. Thus we prove the lemma.


\subsection{Proof of Theorem \ref{thm:neg_pu}\label{sec:proof:thm:neg_pu}}
The proof of Theorem \ref{thm:neg_pu} is a direct formulation of our intuition in Section \ref{sec:neg_pu}. 
Without loss of generality let $(\paper{1}, \ldots, \paper{l})$ be the cycle not reviewed by a single reviewer, for $l\geq 3$. Hence there exists a partial profile $\profile$ such that for all the reviewers who have reviewed both $\paper{\indexpaper}$ and $\paper{j+1}$, $\paper{\indexpaper}\succ \paper{\indexpaper+1}, \forall j\in[l]$ (define $\paper{l+1} = \paper{1}$). On the other hand, since for each reviewer, at least one pair $(\paper{\indexpaper}, \paper{j+1})$ is not reviewed by \him, the constructed partial profile is valid. Now assume $\aggfunctionnotation$ is \PU, then we must have $\paper{1}\succ\cdots \succ \paper{l}$ and $\paper{l} \succ \paper{1}$, which contradicts the transitivity of the ranking. 


\subsection{Proof of Corollary \ref{corol:neg_pu} \label{sec:proof:corol:neg_pu}}

We prove each of the conditions in order. 

~\\\noindent\emph{Proof of part~\ref{point0:neg_pu}:}  If there is a cycle of size $\workload+1$, then no reviewer can review all the papers in it since it exceeds the size of review sets. So there is no such cycle. 

~\\\noindent\emph{Proof of part~\ref{point1:neg_pu}:}  The statement trivially holds for $\workload=2$. For $\workload\geq 3$, Suppose there are two reviewers $\reviewer{\indexreviewer_1}$ and $\reviewer{\indexreviewer_2}$ such that $2\leq |\setpaperselem{\indexreviewer_1}\cap \setpaperselem{\indexreviewer_2}|\leq \workload-1$. Since $\setpaperselem{\indexreviewer_1} \neq \setpaperselem{\indexreviewer_2}$, there exist papers $\paper{\indexpaper_1}$ and $\paper{\indexpaper_2}$ such that $\paper{\indexpaper_1} \in \setpaperselem{\indexreviewer_1} \backslash \setpaperselem{\indexreviewer_2}$ and $\paper{\indexpaper_2} \in \setpaperselem{\indexreviewer_2} \backslash \setpaperselem{\indexreviewer_1}$. Also  $|\setpaperselem{\indexreviewer_1}\cap \setpaperselem{\indexreviewer_2}|\geq 2$, and let $\paper{\indexpaper_3},\paper{\indexpaper_4} \in \setpaperselem{\indexreviewer_1}\cap \setpaperselem{\indexreviewer_2}$. By definition it is easy to verify that $(\paper{\indexpaper_1},\paper{\indexpaper_3},\paper{\indexpaper_2},\paper{\indexpaper_4})$ forms a cycle that satisfies the condition in Theorem \ref{thm:neg_pu}, and hence $(\pairopt)$ is not \PUl.

\newcommand{\paperrelgraph}{\revgraph_p}
~\\\noindent\emph{Proof of part~\ref{point2:neg_pu}:}  Define a ``paper-relation graph'' $\paperrelgraph$ as follows: Given a paper-review assignment $\{\setpaperselem{\indexreviewer}\}_{\indexreviewer=1}^{\numreviewers}$, the paper-relation graph $\paperrelgraph$ is an undirected graph, whose nodes are the \emph{distinct} sets in $\{\setpaperselem{\indexreviewer}\}_{\indexreviewer=1}^{\numreviewers}$; we connect two review sets iff they have one paper in common. Note that by \ref{point1:neg_pu}, each pair of distinct sets has at most one paper in common.

\newcommand{\cycle}{C}
We first show that $(\pairopt)$ is \PUl, then $\paperrelgraph$ must necessarily be a forest. If there is a cycle in $\paperrelgraph$, then there is a corresponding cycle in the review relation graph $\revrelgraph$. To see this, not losing generality suppose the shortest cycle in $\paperrelgraph$ is $(\setpaperselem{1},...,\setpaperselem{l})$. Also, suppose $\setpaperselem{1}\cap \setpaperselem{2}=\{\paper{1}\},\setpaperselem{2}\cap \setpaperselem{3}=\{\paper{2}\}, ..., \setpaperselem{l}\cap \setpaperselem{1}=\{\paper{l}\}$ not losing generality. Then  $(\paper{1},...,\paper{l})$ forms a cycle in $\paperrelgraph$ by its definition. Since each reviewer reviews exactly one set in $\paperrelgraph$, there is no reviewer reviewing all papers in this cycle of papers in $\paperrelgraph$. Thus the condition in Theorem \ref{thm:neg_pu} is satisfied, and $(\pairopt)$ is not \PUl.

\newcommand{\numdistinct}{k_p}
\newcommand{\numedgeprg}{|E_{\paperrelgraph}|}
We now use this result to complete our proof. Consider the union of all sets of papers that form vertices of $\paperrelgraph$. We know that this union contains exactly $\numpapers$ papers since each paper is reviewed at least once. Now let $\numdistinct$ denote the number of distinct review sets (that is, number of vertices of $\paperrelgraph$), and let  $\setpaperselem{\indexreviewer_i},...,\setpaperselem{\indexreviewer_{\numdistinct}}$ denote the vertices of $\paperrelgraph$. The union of three or more sets in $\{\setpaperselem{\indexreviewer_k}\}_{k=1}^{\numdistinct}$ is empty, since otherwise there will be a cycle in $\paperrelgraph$. Using this fact, we apply the inclusion-exclusion principle to obtain
\begin{align*}
    n&=\sum_{k=1}^{\numdistinct} |\setpaperselem{\indexreviewer_k}|-\sum_{1\leq k_1<k_2\leq \numdistinct} |\setpaperselem{\indexreviewer_{k_1}} \cap \setpaperselem{\indexreviewer_{k_2}}|=\numdistinct\workload-\numedgeprg. 
\end{align*}
Now use the inequality $\numedgeprg\leq \numdistinct-1$ which arises since $\paperrelgraph$ is a forest, to obtain the claimed bound $\numdistinct\leq \frac{\numpapers-1}{\workload-1}$.


\subsection{Proof of Proposition \ref{prop:exampleguwsp} \label{sec:proof:prop:exampleguwsp}}

Fix some ranking of papers within each individual set $\tempsetpapers{1}$, $\tempsetpapers{2}$, $\tempsetpapers{3}$ and $\tempsetpapers{4}$ (e.g., according to the natural order of their indices). In the remainder of the proof, any ranking of all papers always considers these fixed rankings within these individual sets. With this in place, in what follows, we refer to any ranking in terms of the rankings of the four sets of papers.

Suppose there is one such $\aggfunctionnotation$ that satisfies group unanimity and weak strategyproofness for $\revgraph$, and consider the following 4 profiles:
\begin{enumerate}[(1)]
    \item $\reviewer{1}:\tempsetpapers{1}\succ \tempsetpapers{2}, \reviewer{2}: \tempsetpapers{2}\succ \tempsetpapers{3}, \reviewer{3}: \tempsetpapers{3}\succ \tempsetpapers{4}$ \label{itm:first_p}
    \inlineitem $\reviewer{1}:\tempsetpapers{2}\succ \tempsetpapers{1}, \reviewer{2}: \tempsetpapers{3}\succ \tempsetpapers{2}, \reviewer{3}: \tempsetpapers{4}\succ \tempsetpapers{3}$\label{itm:seceond_p}
    \item $\reviewer{1}:\tempsetpapers{2}\succ \tempsetpapers{1}, \reviewer{2}: \tempsetpapers{2}\succ \tempsetpapers{3}, \reviewer{3}: \tempsetpapers{3}\succ \tempsetpapers{4}$\label{itm:third_p}
    \inlineitem $\reviewer{1}:\tempsetpapers{2}\succ \tempsetpapers{1}, \reviewer{2}: \tempsetpapers{3}\succ \tempsetpapers{2}, \reviewer{3}: \tempsetpapers{3}\succ \tempsetpapers{4}$\label{itm:fourth_p}
\end{enumerate}
By the property of \GU, profile \ref{itm:first_p} leads to output $\tempsetpapers{1}\succ \tempsetpapers{2}\succ \tempsetpapers{3} \succ \tempsetpapers{4}$, whereas \ref{itm:seceond_p} leads to output $\tempsetpapers{4}\succ \tempsetpapers{3}\succ \tempsetpapers{2} \succ \tempsetpapers{1}$. Now compare \ref{itm:first_p} and \ref{itm:third_p}: The output of \ref{itm:third_p} must have $\tempsetpapers{2}$ at the top and satisfy $\tempsetpapers{3}\succ \tempsetpapers{4}$, by the property of \GU. So the output of profile~\ref{itm:third_p} must be one of i) $\tempsetpapers{2}\succ \tempsetpapers{1}\succ \tempsetpapers{3} \succ \tempsetpapers{4}$, ii) $\tempsetpapers{2}\succ \tempsetpapers{3}\succ \tempsetpapers{1} \succ \tempsetpapers{4}$, or iii) $\tempsetpapers{2}\succ \tempsetpapers{3}\succ \tempsetpapers{4} \succ \tempsetpapers{1}$. Now note that only reviewer $\reviewer{1}$ changes ranking across profiles~\ref{itm:first_p} and~\ref{itm:third_p}, and hence by \WSP the position of at least one paper in the output of profile~\ref{itm:third_p} must be the same as in that of profile~\ref{itm:first_p}. This makes iii) infeasible, so the output of \ref{itm:third_p} must be either i) or ii).  
Similarly, the output of \ref{itm:fourth_p} is either $\tempsetpapers{3}\succ \tempsetpapers{4}\succ \tempsetpapers{2} \succ \tempsetpapers{1}$ or $\tempsetpapers{3}\succ \tempsetpapers{2}\succ \tempsetpapers{4} \succ \tempsetpapers{1}$. Now comparing \ref{itm:third_p} and \ref{itm:fourth_p}: only $\reviewer{2}$ changes ranking, but none of the four papers can be at the same position no matter how we choose the outputs of \ref{itm:third_p} and \ref{itm:fourth_p}. This yields a contradiction.

\newcommand{\numcorres}{\numreviewers'}
\newcommand{\correspaper}{e}
\subsection{Proof of Theorem \ref{thm:neg_pu_total} \label{sec:proof:thm:neg_pu_total}}
\newcommand{\influencegraph}{\revgraph_f}
\newcommand{\edgeinfgraph}{E_{\influencegraph}}

We begin with a definition of an ``influence graph'' $\influencegraph$ induced by any given aggregation rule $\aggfunctionnotation$. 
\begin{definition}[Influence graph]
    \label{def:influence}
    For any review aggregation rule $\aggfunctionnotation$, the influence graph $\influencegraph$ induced by $\aggfunctionnotation$ is a bipartite graph with two groups of vertices $\setreviewers$ and $\setpapers$, and edges as follows. A vertex  $\reviewer{\indexreviewer}$ is connected to vertex $\paper{\indexpaper}$ iff there exists a certain profile $\profile$ such that $\reviewer{\indexreviewer}$ is able to change the output ranking of $\paper{\indexpaper}$ by changing \his own preference. 
    Formally, there exists an edge between any pair  $(\reviewer{\indexreviewer}, \paper{\indexpaper})\in \edgeinfgraph$ iff there exist profiles $\profile = \{\indexedpaperrankshort{1}, \ldots,  \indexedpaperrankshort{i-1}, \indexedpaperrankshort{i}, \indexedpaperrankshort{i+1}, \ldots, \indexedpaperrankshort{m}\}$ and $\profile' = \{\indexedpaperrankshort{1}, \ldots,\indexedpaperrankshort{i-1}, \tilde{\paperrankingnotation}^{(i)},\indexedpaperrankshort{i+1}, \ldots, \indexedpaperrankshort{m}\}$ and $j\in[n]$ such that $\aggfunctionnotation(\profile)(j)\neq \aggfunctionnotation(\profile')(j)$. 
\end{definition}

From this definition, it is thus not hard to see that $\aggfunctionnotation$ is \WSPs if and only if the degree of every reviewer node in $\influencegraph$ is strictly smaller than $\numpapers$.

We prove the claim via a contradiction argument. Assume that $\aggfunctionnotation$ is both \PUs and \WSPs. Let $\influencegraph$ be the corresponding influence graph. 
Firstly we show that $\deg(\genpaper)>0$ for every paper $\genpaper$, where the degree is for the influence graph $\influencegraph$. Suppose otherwise that $\deg(\paper{\indexpaper})=0$ for some paper $\paper{\indexpaper}$. This means \emph{no} reviewer can affect the ranking of $\paper{\indexpaper}$; in other words, the position of paper $\paper{\indexpaper}$ is fixed regardless of the profile. This contradicts with our assumption of \PUln; to see this, pick another paper $\paper{\indexpaper'}$ where $\indexpaper'\ne \indexpaper$ (this is possible since $\numpapers\geq 2$). Not losing generality suppose $\indexpaper< \indexpaper'$. Consider a profile $\profile$ where every reviewer ranks $\paper{\indexpaper}\succ \paper{\indexpaper'} \succ \tempsetpapers{-(\indexpaper,\indexpaper')}$, and another profile $\profile'$ where everyone ranks $\paper{\indexpaper‘}\succ \paper{\indexpaper} \succ \tempsetpapers{-(\indexpaper,\indexpaper')}$; here $\tempsetpapers{-(\indexpaper,\indexpaper')}$ means the ordinal ranking of papers other than $\paper{\indexpaper},\paper{\indexpaper'}$, i.e., $\paper{1}\succ\cdots\succ \paper{\indexpaper-1}\succ \paper{\indexpaper+1}\succ \cdots\succ \paper{\indexpaper'-1}\succ \paper{\indexpaper'+1}\succ\cdots\succ \paper{\numpapers}$. By the property of \PU, when everyone ranks the same the final result must be the same as everyone; however this means the position of $\paper{\indexpaper}$ is different in the two profiles, and thus the position of $\paper{\indexpaper}$ is not fixed. This makes contradiction and we prove that $\deg(\genpaper)> 0$ for every paper $\genpaper$.

Now for any reviewer $\reviewer{\indexreviewer}\in \setreviewers$, let $\correspaper(\reviewer{\indexreviewer})\in \setpapers$ be the paper with the lowest index in $\setpapers$  such that $(\reviewer{\indexreviewer}, \correspaper(\reviewer{\indexreviewer}))\not\in \edgeinfgraph$. Since $\aggfunctionnotation$ is \WSPs, $\correspaper(\reviewer{\indexreviewer})$ must exist for all $\indexreviewer\in[\numreviewers]$. Define the set of such papers as $\setpapers' \defeq \{\correspaper_1, \ldots, \correspaper_{\numcorres}\} = \{\correspaper(\reviewer{\indexreviewer}): \reviewer{\indexreviewer}\in \setreviewers\}$. Note that we must have $\numcorres\leq \numreviewers$ and in fact $\numcorres$ can be strictly smaller than $m$ because of the possible overlap between $\correspaper(\reviewer{\indexreviewer}), \forall \indexreviewer\in[\numreviewers]$. From the definition of $\numcorres$ and property of \WSP, it is clear that $\numcorres\geq 1$. If $\numcorres=1$, we have $(\reviewer{\indexreviewer},\correspaper_1)\not\in \edgeinfgraph$ for every reviewer $\reviewer{i}$; this contradicts with the fact that $\deg(\genpaper)\geq 0$ for every paper $\genpaper$. So $\numcorres>1.$ 

In this proof, we slightly overload the notation of $\aggfunctionnotation_{\correspaper_k}(\profile)$ to mean the position of paper $\correspaper_k$ in $\aggfunctionnotation(\profile)$.
Based on the inverse mapping from $\setpapers'$ to $\setreviewers$, we partition all the reviewers $\setreviewers$ into $\numcorres$ groups $\{\setreviewers'_1, \ldots, \setreviewers'_{\numcorres}\}$ such that all reviewers in any set $\setreviewers'_k$ contributed paper $\correspaper_k$ when defining set $\setpapers'$. In particular, we have that no reviewer in $\setreviewers'_j$  is connected to paper $\correspaper_j$ in the influence graph $\influencegraph$. 

In the description that follows, we restrict attention to the papers in $\setpapers'$, and assume that in any ranking all remaining $(\numpapers - \numpapers')$ papers are positioned at the end of the preference list of any reviewer. Now consider the following two preferences over $\setpapers'$:
\begin{align*}
\paperrankingnotation = \correspaper_1\succ \correspaper_2\succ \cdots \succ \correspaper_{\numcorres}, \quad \paperrankingnotation' = \correspaper_2\succ \correspaper_3\succ\cdots \succ \correspaper_{\numcorres}\succ \correspaper_1.
\end{align*}
Using $\paperrankingnotation$ and $\paperrankingnotation'$, define the following $\numcorres+1$ different profiles:
\begin{align*}
\profile_0 = (\paperrankingnotation, \ldots, \paperrankingnotation), ~ \profile_k = (\underbrace{\paperrankingnotation', \ldots, \paperrankingnotation'}_{k}, \underbrace{\paperrankingnotation, \ldots, \paperrankingnotation}_{\numcorres-k}) ~~\forall k\in[\numcorres].
\end{align*}
where in $\profile_k$ the first $k$ preferences are $\paperrankingnotation'$ and the last ($\numcorres-k$)  preferences are $\paperrankingnotation$. In what follows, we will use a diagonalization argument to generate a contradiction using the condition that $\aggfunctionnotation$ is \WSPs. We first present a lemma, which we prove in Section~\ref{sec:prooflemdiagonal}.
\begin{lemma}
    \label{lemma:diagonal}
    If $\aggfunctionnotation$ is \PUl and weakly strategyproof, $\aggfunctionnotation_{\correspaper_k}(\profile_k) = k $ for every $k \in [\numcorres]$, that is, under profile $\profile_k$, the $k$-th position in the output ranking must be taken by paper $\correspaper_k$.
\end{lemma}

Applying Lemma \ref{lemma:diagonal} with $k=\numcorres$, we obtain that $\aggfunctionnotation_{\correspaper_{\numreviewers'}}(\profile_{\numreviewers'}) = {\numreviewers'}$. However, on the other side, since $\profile_{\numcorres} = (\paperrankingnotation', \ldots, \paperrankingnotation')$ and $\paperrankingnotation' = \correspaper_2\succ \correspaper_3\succ\cdots \succ \correspaper_{\numcorres}\succ \correspaper_1$, again by the \PUs property we have $\aggfunctionnotation_{\correspaper_{\numreviewers'}}(\profile_{\numreviewers'}) = 1$. This leads to a contradiction, hence $\aggfunctionnotation$ cannot be both \WSPs and \PUs.

\subsubsection{Proof of Lemma~\ref{lemma:diagonal}}
\label{sec:prooflemdiagonal}
We prove by induction on $k$. 

\noindent\textbf{Base case}. Since $\aggfunctionnotation$ is \PUs, the output ranking $\aggfunctionnotation(\profile_{0})$ must be:
$$\aggfunctionnotation(\profile_{0}) = \correspaper_1\succ \correspaper_2\succ\cdots \succ \correspaper_{\numcorres}$$
Consider $k = 1$. Note that $\paperrankingnotation$ and $\paperrankingnotation'$ differ only at the position of $\correspaper_1$, and in $\profile_{1}$, only $\setpapers'_1$ changes their preference and all the other preferences are kept fixed. Then by the \WSPs of $\aggfunctionnotation$, the output ranking of $\correspaper_1$ will not be changed because $\setpapers'_1$ are not connected to $\correspaper_1$ in the influence graph, so we must have:
\begin{align*}
    \aggfunctionnotation(\profile_{1}) = \correspaper_1\succ \correspaper_2\succ\cdots \succ \correspaper_{\numcorres},
\end{align*}
\textbf{Induction step}. Suppose the claim of this lemma holds for $\{1,\ldots,k\}$. Consider the case of $k+1$. 

Observe that $\aggfunctionnotation(\profile_k)(\correspaper_k) = k$, and in both $\paperrankingnotation$ and $\paperrankingnotation'$ we have:
\begin{align*}
    \correspaper_k\succ \correspaper_{k+1}\succ\cdots \succ \correspaper_{\numcorres}. 
\end{align*}
Then since $\aggfunctionnotation$ is PU, we know that the last $\numcorres-k+1$ positions in the output ranking of $\aggfunctionnotation(\profile_k)$ must be given by $\correspaper_k\succ \correspaper_{k+1}\succ\cdots \succ \correspaper_{\numcorres}$, i.e., $\aggfunctionnotation_{\correspaper_{k+1}}(\profile_k) = k+1$. The profiles $\profile_k$ and $\profile_{k+1}$ differ only in the preference given by $\setpapers'_{k+1}$, and no reviewer in set $\setpapers'_{k+1}$ can influence the position of paper $\correspaper_{k+1}$. It follows that $\aggfunctionnotation_{\correspaper_{k+1}}(\profile_{k+1}) = k+1$, which completes our proof.

\section{\label{sec:conclusion}Discussion}
In this paper we address the problem of designing strategyproof and efficient peer-review mechanism. The setting of peer review is challenging due to the various idiosyncrasies of the peer-review process: reviewers review only a subset of papers, each paper has multiple authors who may be reviewers, and each reviewer may author multiple submissions. We provide a framework and associated algorithms to impart strategyproofness to conference peer review. Our framework, besides guaranteeing strategyproofness, is importantly very flexible in allowing the program chairs to use the decision-making criteria of their choice. We complement these positive results with negative results showing that it is impossible for any algorithm to remain strategyproof and satisfy the stronger notion of pairwise unanimity. Future work includes considering efficiency from a statistical perspective and characterizing the precise set of conflict-of-interest graphs that permit (or not) strategyproofness.


The framework established here leads to a number of useful open problems:
\begin{itemize}
\item Can recruitment of a small number of reviewers with no conflicts (e.g., in case of authorship conflicts, reviewers who have not submitted any papers) lead to significant improvements in efficiency? Can better ways to eliminate some authors from the reviewer pool increase applicability of partition-based algorithms?
\item The results in this paper considered the social choice property of unanimity as a measure of efficiency. While this can be regarded as a first-order notion of efficiency, it is of interest to consider complex notions of efficiency. One useful notion of efficiency is the statistical utility of estimation~\citep{stelmakh2018assignment} of the (partial or full) ranking of papers under a statistical model for reviewer reports. An alternative notion of efficiency combines an assignment quality based on the similarities of assigned reviewers and papers with group unanimity (e.g., maximizing the similarity scores in the assignment while preserving group unanimity).

\end{itemize}


\section*{Acknowledgments}
This work was supported in parts by NSF grants CRII: CIF: 1755656 and CIF: 1763734.


\bibliography{reference}

\begin{thebibliography}{}

\bibitem[Alon et~al., 2011]{alon2011sum}
Alon, N., Fischer, F., Procaccia, A., and Tennenholtz, M. (2011).
\newblock Sum of us: Strategyproof selection from the selectors.
\newblock In {\em Proceedings of the 13th Conference on Theoretical Aspects of
  Rationality and Knowledge}, pages 101--110. ACM.

\bibitem[Anderson et~al., 2007]{anderson2007perverse}
Anderson, M.~S., Ronning, E.~A., De~Vries, R., and Martinson, B.~C. (2007).
\newblock The perverse effects of competition on scientists’ work and
  relationships.
\newblock {\em Science and engineering ethics}, 13(4):437--461.

\bibitem[Arrow, 1950]{arrow1950difficulty}
Arrow, K.~J. (1950).
\newblock A difficulty in the concept of social welfare.
\newblock {\em Journal of political economy}, 58(4):328--346.

\bibitem[Aziz et~al., 2016]{aziz2016strategyproof}
Aziz, H., Lev, O., Mattei, N., Rosenschein, J.~S., and Walsh, T. (2016).
\newblock Strategyproof peer selection: Mechanisms, analyses, and experiments.
\newblock In {\em AAAI}, pages 397--403.

\bibitem[Aziz et~al., 2019]{aziz2019strategyproof}
Aziz, H., Lev, O., Mattei, N., Rosenschein, J.~S., and Walsh, T. (2019).
\newblock Strategyproof peer selection using randomization, partitioning, and
  apportionment.
\newblock {\em Artificial Intelligence}.

\bibitem[Balietti et~al., 2016]{balietti2016peer}
Balietti, S., Goldstone, R.~L., and Helbing, D. (2016).
\newblock Peer review and competition in the art exhibition game.
\newblock {\em Proceedings of the National Academy of Sciences},
  113(30):8414--8419.

\bibitem[Barnett, 2003]{barnett2003modern}
Barnett, W. (2003).
\newblock The modern theory of consumer behavior: Ordinal or cardinal?
\newblock {\em The Quarterly Journal of Austrian Economics}, 6(1):41--65.

\bibitem[Bird et~al., 2009]{bird2009natural}
Bird, S., Loper, E., and Klein, E. (2009).
\newblock Natural language processing with python o’reilly media inc.

\bibitem[Bousquet et~al., 2014]{bousquet2014near}
Bousquet, N., Norin, S., and Vetta, A. (2014).
\newblock A near-optimal mechanism for impartial selection.
\newblock In {\em International Conference on Web and Internet Economics},
  pages 133--146. Springer.

\bibitem[Brandt et~al., 2016]{brandt2016handbook}
Brandt, F., Conitzer, V., Endriss, U., Procaccia, A.~D., and Lang, J. (2016).
\newblock {\em Handbook of computational social choice}.
\newblock Cambridge University Press.

\bibitem[Caragiannis et~al., 2017]{caragiannis2017optimizing}
Caragiannis, I., Chatzigeorgiou, X., Krimpas, G.~A., and Voudouris, A.~A.
  (2017).
\newblock Optimizing positional scoring rules for rank aggregation.
\newblock In {\em AAAI}, pages 430--436.

\bibitem[Charlin and Zemel, 2013]{charlin13tpms}
Charlin, L. and Zemel, R.~S. (2013).
\newblock The {T}oronto {P}aper {M}atching {S}ystem: An automated
  paper-reviewer assignment system.

\bibitem[Connolly et~al., 2014]{connolly2014longitudinal}
Connolly, R., Miller, J., and Friedman, R. (2014).
\newblock A longitudinal examination of sigite conference submission data,
  2007-2012.
\newblock In {\em Proceedings of the 15th Annual Conference on Information
  technology education}, pages 167--172. ACM.

\bibitem[De~Clippel et~al., 2008]{de2008impartial}
De~Clippel, G., Moulin, H., and Tideman, N. (2008).
\newblock Impartial division of a dollar.
\newblock {\em Journal of Economic Theory}, 139(1):176--191.

\bibitem[D{\'\i}ez~Pel{\'a}ez et~al., 2013]{diez2013peer}
D{\'\i}ez~Pel{\'a}ez, J., Luaces~Rodr{\'\i}guez, {\'O}., Alonso~Betanzos, A.,
  Troncoso, A., and Bahamonde~Rionda, A. (2013).
\newblock Peer assessment in moocs using preference learning via matrix
  factorization.
\newblock In {\em NIPS Workshop on Data Driven Education}.

\bibitem[D{\"o}rfler et~al., 2017]{dorfler2017incentive}
D{\"o}rfler, F., Xiao, Y., and van~der Schaar, M. (2017).
\newblock Incentive design in peer review: Rating and repeated endogenous
  matching.
\newblock {\em IEEE Transactions on Network Science and Engineering}.

\bibitem[Douceur, 2009]{douceur2009paper}
Douceur, J.~R. (2009).
\newblock Paper rating vs. paper ranking.
\newblock {\em ACM SIGOPS Operating Systems Review}, 43(2):117--121.

\bibitem[Emerson, 2013]{emerson2013original}
Emerson, P. (2013).
\newblock {The original Borda count and partial voting}.
\newblock {\em Social Choice and Welfare}, pages 1--6.

\bibitem[Fiez et~al., 2019]{fiez2019super}
Fiez, T., Shah, N., and Ratliff, L. (2019).
\newblock A {SUPER}* algorithm to optimize paper bidding in peer review.
\newblock In {\em ICML workshop on Real-world Sequential Decision Making:
  Reinforcement Learning And Beyond}.

\bibitem[Fischer and Klimm, 2015]{fischer2015optimal}
Fischer, F. and Klimm, M. (2015).
\newblock Optimal impartial selection.
\newblock {\em SIAM Journal on Computing}, 44(5):1263--1285.

\bibitem[Fishburn, 2015]{fishburn2015theory}
Fishburn, P.~C. (2015).
\newblock {\em The theory of social choice}.
\newblock Princeton University Press.

\bibitem[Fulkerson and Gross, 1965]{fulkerson1965incidence}
Fulkerson, D. and Gross, O. (1965).
\newblock Incidence matrices and interval graphs.
\newblock {\em Pacific journal of mathematics}, 15(3):835--855.

\bibitem[Gao et~al., 2019]{gao2019does}
Gao, Y., Eger, S., Kuznetsov, I., Gurevych, I., and Miyao, Y. (2019).
\newblock Does my rebuttal matter? insights from a major nlp conference.
\newblock {\em arXiv preprint arXiv:1903.11367}.

\bibitem[Garg et~al., 2010]{Garg2010papers}
Garg, N., Kavitha, T., Kumar, A., Mehlhorn, K., and Mestre, J. (2010).
\newblock Assigning papers to referees.
\newblock {\em Algorithmica}, 58(1):119--136.

\bibitem[Ge et~al., 2013]{ge13bias}
Ge, H., Welling, M., and Ghahramani, Z. (2013).
\newblock A {B}ayesian model for calibrating conference review scores.

\bibitem[Gibbard, 1973]{gibbard1973manipulation}
Gibbard, A. (1973).
\newblock Manipulation of voting schemes: a general result.
\newblock {\em Econometrica: journal of the Econometric Society}, pages
  587--601.

\bibitem[Hajek et~al., 2014]{hajek2014minimax}
Hajek, B., Oh, S., and Xu, J. (2014).
\newblock Minimax-optimal inference from partial rankings.
\newblock In {\em Advances in Neural Information Processing Systems}, pages
  1475--1483.

\bibitem[Hartvigsen et~al., 1999]{Hartvigsen99assignment}
Hartvigsen, D., Wei, J.~C., and Czuchlewski, R. (1999).
\newblock The conference paper-reviewer assignment problem.
\newblock {\em Decision Sciences}, 30(3):865--876.

\bibitem[Hazelrigg, 2013]{hazelrigg2013dear}
Hazelrigg, G. (2013).
\newblock {Dear colleague letter: Information to principal investigators (PIs)
  planning to submit proposals to the Sensors and Sensing Systems (SSS) program
  October 1, 2013, deadline}.
\newblock {\em Deadline (NSF Website,
  http://www.nsf.gov/pubs/2013/nsf13096/nsf13096.jsp)}.

\bibitem[Hojat et~al., 2003]{hojat2003impartial}
Hojat, M., Gonnella, J.~S., and Caelleigh, A.~S. (2003).
\newblock Impartial judgment by the “gatekeepers” of science: fallibility
  and accountability in the peer review process.
\newblock {\em Advances in Health Sciences Education}, 8(1):75--96.

\bibitem[Holzman and Moulin, 2013]{holzman2013impartial}
Holzman, R. and Moulin, H. (2013).
\newblock Impartial nominations for a prize.
\newblock {\em Econometrica}, 81(1):173--196.

\bibitem[Kahng et~al., 2017]{kahng2017ranking}
Kahng, A.~B., Kotturi, Y., Kulkarni, C., Kurokawa, D., and Procaccia, A.~D.
  (2017).
\newblock Ranking wily people who rank each other.
\newblock {\em Technical Report}.

\bibitem[Kurokawa et~al., 2015]{kurokawa2015impartial}
Kurokawa, D., Lev, O., Morgenstern, J., and Procaccia, A.~D. (2015).
\newblock Impartial peer review.
\newblock In {\em IJCAI}, pages 582--588.

\bibitem[Langford, 2008]{langford201adversarialacad}
Langford, J. (2008).
\newblock Adversarial academia.
\newblock \url{http://hunch.net/?p=499}.

\bibitem[Lawrence and Cortes, 2014]{nips14experiment}
Lawrence, N. and Cortes, C. (2014).
\newblock {The NIPS Experiment}.
\newblock \url{http://inverseprobability.com/2014/12/16/the-nips-experiment}.
\newblock [Online; accessed 3-June-2017].

\bibitem[Makhorin, 2001]{makhorin2001gnu}
Makhorin, A. (2001).
\newblock Gnu linear programming kit.
\newblock {\em Moscow Aviation Institute, Moscow, Russia}, 38.

\bibitem[Mathieus, 2008]{claire2008soda}
Mathieus, C. (2008).
\newblock {SODA PC} meetings.
\newblock \url{http://cs.brown.edu/~claire/SODAnotes.pdf} (last retrieved May
  21, 2018).

\bibitem[Merrifield and Saari, 2009]{merrifield2009telescope}
Merrifield, M.~R. and Saari, D.~G. (2009).
\newblock Telescope time without tears: a distributed approach to peer review.
\newblock {\em Astronomy \& Geophysics}, 50(4):4--16.

\bibitem[Noothigattu et~al., 2018]{noothigattu2018choosing}
Noothigattu, R., Shah, N., and Procaccia, A. (2018).
\newblock Choosing how to choose papers.
\newblock {\em arXiv preprint arxiv:1808.09057}.

\bibitem[Piech et~al., 2013]{piech2013tuned}
Piech, C., Huang, J., Chen, Z., Do, C., Ng, A., and Koller, D. (2013).
\newblock Tuned models of peer assessment in moocs.
\newblock {\em arXiv preprint arXiv:1307.2579}.

\bibitem[Roos et~al., 2011]{roos2011calibrate}
Roos, M., Rothe, J., and Scheuermann, B. (2011).
\newblock How to calibrate the scores of biased reviewers by quadratic
  programming.
\newblock In {\em AAAI}.

\bibitem[Salton et~al., 1975]{salton1975vector}
Salton, G., Wong, A., and Yang, C.-S. (1975).
\newblock A vector space model for automatic indexing.
\newblock {\em Communications of the ACM}, 18(11):613--620.

\bibitem[Satterthwaite, 1975]{satterthwaite1975strategy}
Satterthwaite, M.~A. (1975).
\newblock Strategy-proofness and arrow's conditions: Existence and
  correspondence theorems for voting procedures and social welfare functions.
\newblock {\em Journal of economic theory}, 10(2):187--217.

\bibitem[Sch{\"u}tze et~al., 2008]{schutze2008introduction}
Sch{\"u}tze, H., Manning, C.~D., and Raghavan, P. (2008).
\newblock Introduction to information retrieval.
\newblock In {\em Proceedings of the international communication of association
  for computing machinery conference}, volume~4.

\bibitem[Shah et~al., 2016]{shah2016estimation}
Shah, N.~B., Balakrishnan, S., Bradley, J., Parekh, A., Ramchandran, K., and
  Wainwright, M.~J. (2016).
\newblock Estimation from pairwise comparisons: Sharp minimax bounds with
  topology dependence.
\newblock {\em The Journal of Machine Learning Research}, 17(1):2049--2095.

\bibitem[Shah et~al., 2013]{shah2013case}
Shah, N.~B., Bradley, J.~K., Parekh, A., Wainwright, M., and Ramchandran, K.
  (2013).
\newblock A case for ordinal peer-evaluation in moocs.
\newblock In {\em {NIPS Workshop on Data Driven Education}}.

\bibitem[Shah et~al., 2017]{shah2017design}
Shah, N.~B., Tabibian, B., Muandet, K., Guyon, I., and von Luxburg, U. (2017).
\newblock {Design and Analysis of the NIPS 2016 Review Process}.
\newblock {\em arXiv preprint arXiv:1708.09794}.

\bibitem[Stelmakh et~al., 2019a]{stelmakh2019testing}
Stelmakh, I., Shah, N., and Singh, A. (2019a).
\newblock On testing for biases in peer review.
\newblock In {\em NeurIPS}.

\bibitem[Stelmakh et~al., 2019b]{stelmakh2018assignment}
Stelmakh, I., Shah, N.~B., and Singh, A. (2019b).
\newblock {PeerReview4All}: Fair and accurate reviewer assignment in peer
  review.
\newblock In {\em Algorithmic Learning Theory}.

\bibitem[Stewart et~al., 2005]{stewart2005absolute}
Stewart, N., Brown, G.~D., and Chater, N. (2005).
\newblock Absolute identification by relative judgment.
\newblock {\em Psychological review}, 112(4):881.

\bibitem[Thurner and Hanel, 2011]{thurner2011peer}
Thurner, S. and Hanel, R. (2011).
\newblock Peer-review in a world with rational scientists: Toward selection of
  the average.
\newblock {\em The European Physical Journal B}, 84(4):707--711.

\bibitem[Tomkins et~al., 2017]{tomkins2017reviewer}
Tomkins, A., Zhang, M., and Heavlin, W.~D. (2017).
\newblock Reviewer bias in single-versus double-blind peer review.
\newblock {\em Proceedings of the National Academy of Sciences},
  114(48):12708--12713.

\bibitem[Tsukida and Gupta, 2011]{tsukida2011analyze}
Tsukida, K. and Gupta, M.~R. (2011).
\newblock How to analyze paired comparison data.
\newblock Technical report, DTIC Document.

\bibitem[Wang and Shah, 2019]{wang2018your}
Wang, J. and Shah, N.~B. (2019).
\newblock Your 2 is my 1, your 3 is my 9: Handling arbitrary miscalibrations in
  ratings.
\newblock In {\em AAMAS}.

\end{thebibliography}

\appendix
\section{More Details on the ICLR 2018 Experiment}
\label{sec:appexp}
In this section we describe in more detail on the similar scoring model and the optimization formulation used in our ICLR 2018 experiment in Section~\ref{sec:analy_iclr_2018}.

\subsection{Similarity Scoring Model}
We first use text representations to model the reviewers. In particular, for each reviewer in the pool, we scrape their (at most 10) most recent papers from \href{https://arxiv.org/}{arXiv}\footnote{To ensure that the downloaded papers belong to the corresponding author in the general area of artificial intelligence, we only scrape papers under the following categories: cs.LG, cs.AI, stat.ML, cs.CV, cs.NE, cs.CL, cs.GT and cs.RO.} as the corresponding text representation. To preprocess the text of papers and reviewers, we remove the stop words, tokenize the text, and then use the \texttt{PorterStemmer}~\citep{bird2009natural} to obtain the word stem of each word. After the preprocessing step, the dictionary contains $d = 290158$ unique words. Based on this dictionary, we use the vector space model~\citep{salton1975vector} to represent each reviewer/paper as a vector in $\RR^d$. The ICLR 2018 data contains 911 submitted papers and 2435 reviewers, hence there are $N = 3346$ documents in total.

To compute the similarity scores between reviewers and papers, for each document $D$ (reviewer or paper), we compute the corresponding term frequency-inverse document frequency (tf-idf)~\citep{schutze2008introduction} score as the vector representation. Specifically, for a term $w$ in the document, we use $N_w$ to denote the number of times $w$ appears in the corpus that contains $N$ documents. Then the inverse document frequency of the term $w$ is given by:
\begin{equation*}
    \idf(w)\defeq \log\frac{N}{N_w}.
\end{equation*}
To prevent a bias towards longer documents, e.g., raw frequency of $w$ divided by the raw frequency of the most occurring term in the document $D$, we use the following augmented frequency as the term frequency of $w$ in document $D$:
\begin{equation*}
    \tf(w, D)\defeq \frac{1}{2} + \frac{1}{2}\frac{f_{w, D}}{\max\{f_{w', D}:w'\in D\}},
\end{equation*}
where we use $f_{w,D}$ to denote the number of times term $w$ appearing in $D$. Let $v_D\in\RR^d$ be the vector representation of document $D$. Then the value of the coordinate corresponding to term $w$ is given by:
\begin{equation*}
    v_D(w) = \tf(w, D)\times \idf(w) = \left(\frac{1}{2} + \frac{1}{2}\frac{f_{w, D}}{\max\{f_{w', D}:w'\in D\}}\right) \times \log\frac{N}{N_w}.
\end{equation*}
We then construct the similarity matrix $S\in\RR^{\numreviewers\times \numpapers}$ between reviewers and papers whose each entry $\simscore_{ij}\in [0, 1]$ corresponds to the similarity score between reviewer $\reviewer{i}$ and paper $\paper{j}$. $\simscore_{ij}$ is given by the cosine similarity of the corresponding tf-idf vectors:
\begin{equation*}
    \simscore_{ij} = \frac{v_{\reviewer{i}}^T v_{\paper{j}}}{\|v_{\reviewer{i}}\|_2\cdot\|v_{\paper{j}}\|_2}\in[0, 1].
\end{equation*}

\subsection{The Reviewer-Paper Assignment Algorithm \label{app:assignment_tpms}}
Matching is the process of assigning papers to reviewers. Given the similarity score matrix $S\in\RR^{\numreviewers\times\numpapers}$, we solve the following optimization problem, as used in the current TPMS system, to compute the assignment. The optimization problem formulated in \eqref{equ:assopt} is an integer program, where the objective function corresponds to maximizing the sum of similarity scores in the matching. Here for any reviewer-paper pair $(i, j)$, we have $a_{ij} = 1$ iff paper $\paper{j}$ is assigned to reviewer $\reviewer{i}$ in the matching:
\begin{equation}
    \begin{aligned}
    & \underset{a_{ij}}{\text{maximize}} && \sum_{i\in[\numreviewers]}\sum_{j\in[\numpapers]} \simscore_{ij} a_{ij} \\
    & \text{subject to} && a_{ij}\in \{0, 1\}, \quad\forall i\in[\numreviewers], j\in[\numpapers] \\
    &                   && \sum_{j}a_{ij}\leq\maxpapers, \quad \forall i\in[\numreviewers]\\
    &                   && \sum_{i}a_{ij} \geq \minreviewers,\quad\forall j\in[\numpapers]
    \end{aligned}
    \label{equ:assopt}
\end{equation}
The constraint $\sum_{j}a_{ij}\leq\maxpapers$ means that we restrict the maximum number of papers assigned to a reviewer to be $\maxpapers$. Furthermore, we also use the constraint $\sum_{i}a_{ij} \geq \minreviewers$ to enforce that each paper should be reviewed by at least $\minreviewers$ reviewers. In the ICLR 2018 data the number of optimization variables $a_{ij}$ is more than 2 million, which is intractable to solve using existing integer program solvers. So instead, we can relax the above integer program to the following linear program (LP):
\begin{equation}
    \begin{aligned}
    & \underset{a_{ij}}{\text{maximize}} && \sum_{i\in[\numreviewers]}\sum_{j\in[\numpapers]} \simscore_{ij} a_{ij} \\
    & \text{subject to} && 0 \leq a_{ij} \leq 1,  \quad \forall i\in[\numreviewers], j\in[\numpapers] \\
    &                   && \sum_{j}a_{ij}\leq\maxpapers, \quad \forall i\in[\numreviewers]\\
    &                   && \sum_{i}a_{ij} \geq \minreviewers,\quad\forall j\in[\numpapers]
    \end{aligned}
    \label{equ:lpopt}
\end{equation}
In the above LP we relax the integral constraint over $a_{ij}$ in~\eqref{equ:assopt} to $0\leq a_{ij}\leq 1$, $\forall i \in[\numreviewers],j\in[\numpapers]$. Due to the relaxation, it is clear that the optimal value of~\eqref{equ:lpopt} is at least that of~\eqref{equ:assopt}. On the other hand, observe that if we reformulate the constraints $\sum_{j}a_{ij}\leq\maxpapers$ and $\sum_{i}a_{ij} \geq \minreviewers$ into the matrix form, then the corresponding constraint matrix will be the node-edge incidence matrix of a complete bipartite graph consisting of a set of reviewers and a set of papers. It follows from a known sufficient condition~\citep{fulkerson1965incidence} that the incidence matrix of a bipartite graph is \emph{totally unimodular}, which implies that the solution of the LP in~\eqref{equ:lpopt} is guaranteed to be integral. Hence in order to obtain the optimal solution of~\eqref{equ:assopt}, we can use existing polynomial time solvers to solve the relaxed LP, and since the constraint matrix in~\eqref{equ:assopt} is totally unimodular, this gives us a polynomial time algorithm to compute the optimal solution of~\eqref{equ:assopt}. In our implementation we use the GNU Linear Programming Kit (GLPK)~\citep{makhorin2001gnu} that implements the simplex algorithm to solve the LP in~\eqref{equ:lpopt}.

\end{document}